\documentclass[]{acmsiggraph}

\pdfoutput=1

\newcommand{\mytitle}{Interpolations of Smoke and Liquid Simulations}

\usepackage{bm}  %
\usepackage{algpseudocode}
\usepackage{amsmath}
\usepackage{amssymb}
\usepackage{comment} 
\usepackage{algorithm2eNT}

\usepackage{color}
\definecolor{green}{rgb}{0.125,0.65,0.125}
\definecolor{reddy}{rgb}{0.7,0.,0.}
\definecolor{greeny}{rgb}{0.1,0.7,0.}
\definecolor{orange}{RGB}{255,74,13}
\definecolor{water}{rgb}{0.25,0.35,0.75}

\newcommand{\revision}[1]{#1}
\newcommand{\scndRev}[1]{#1}
\newcommand{\thirdRev}[1]{#1}
\newcommand{\fourthRev}[1]{#1}

\newcommand{\mypara}[1]{{\bf #1 \ }}
\newcommand{\mycdt}{ \hspace{-1.1pt} \cdot  \hspace{-1.1pt} }

\newcommand{\vspaceImgSm}{ \vspace{-0.085cm} }
\newcommand{\vspaceImgMed}{ \vspace{-0.12cm} }

\newcommand{\myrefeq}[1]{Eq.~(\ref{#1})}
\newcommand{\myreffig}[1]{Figure~\ref{#1}}
\newcommand{\myreftab}[1]{Table~\ref{#1}}
\newcommand{\myrefsec}[1]{Section~\ref{#1}}

\newcommand{\myrefalg}[1]{Algorithm~\ref{#1}}

\newcommand{\Vect}[1]{{\bf{#1}}}
\newcommand{\Mat}[1]{{\bf{#1}}}

\newcommand{\veldst}{{\velV_{\text{comb}}}}

\newcommand{\velVc}{u}
\newcommand{\imVc}{\Phi} 

\newcommand{\velV}{{\bf{u}}}
\newcommand{\imV}{\boldsymbol{\Phi}}

\newcommand{\GradPhiA}{{\boldsymbol{[\nabla\Phi_2]}}}

\newcommand{\uvfunc}{H}
\newcommand{\uvp}{\mathbf{x}}
\newcommand{\uvpt}{\tilde{\mathbf{x}}}
\newcommand{\imVal}{\Phi} 
\newcommand{\imDstal}{{\Phi_{\text{dst}}}}
\newcommand{\velAl}{u}
\newcommand{\posAl}{x}
\newcommand{\matof}{\Mat{A}_{of}}
\newcommand{\imVbl}{\boldsymbol{\Psi}} 

\newcommand{\defo}[2]{\mathbf{u}_{#1 \rightarrow #2}}
\newcommand{\dproj}{\tau_{\text{proj}}}

\newlength{\punctuationfootlength}

\author{Nils Thuerey, \ \ \  {\em Technical University of Munich} }
\pdfauthor{Nils Thuerey} %

\begin{document}

\title{\mytitle}

\maketitle

\begin{abstract}

We present a novel method to interpolate smoke and liquid simulations in order to perform data-driven fluid simulations.  Our approach calculates a dense space-time deformation using grid-based signed-distance functions of the inputs.

A key advantage of this implicit Eulerian representation is that it allows us to use powerful techniques from the optical flow area. We employ a five-dimensional optical flow solve. In combination with a projection algorithm, and residual iterations, we achieve a robust matching of the inputs. Once the match is computed, arbitrary in between variants can be created very efficiently. To concatenate multiple long-range deformations, we propose a novel alignment technique.

Our approach has numerous advantages, including automatic matches without user input, volumetric deformations that can be applied to details around the surface, and the inherent handling of topology changes. As a result, we can interpolate swirling smoke clouds, and splashing liquid simulations.  We can even match and interpolate phenomena with fundamentally different physics: a drop of liquid, and a blob of heavy smoke.

\end{abstract}

\begin{figure}[ht!]
   \vspace{-0.2cm} \begin{center}
	\includegraphics[width=0.490\textwidth]{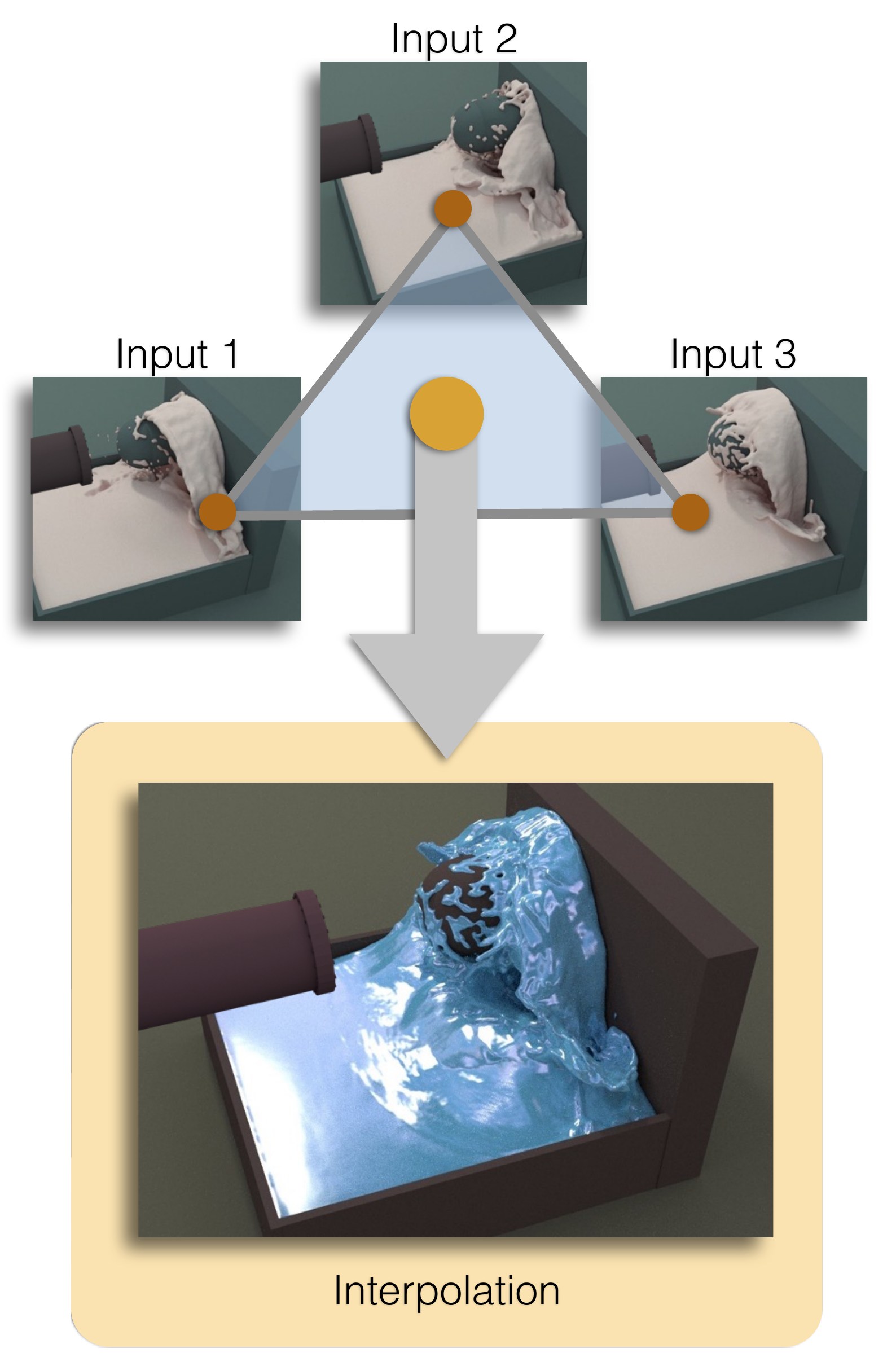} \end{center}  \vspaceImgSm
   \caption{ \label{fig:teaser}
	Our method calculates matches between 4D data sets of smoke and liquid simulations, which
	are then used to calculate arbitrary in-betweens. A liquid example with a 2D parameter space
	is shown above. }
\end{figure}

\section{Introduction}
\label{sec:intro}
\begin{figure*}[t]
{ \begin{center}
	\includegraphics[width=1.0\textwidth]{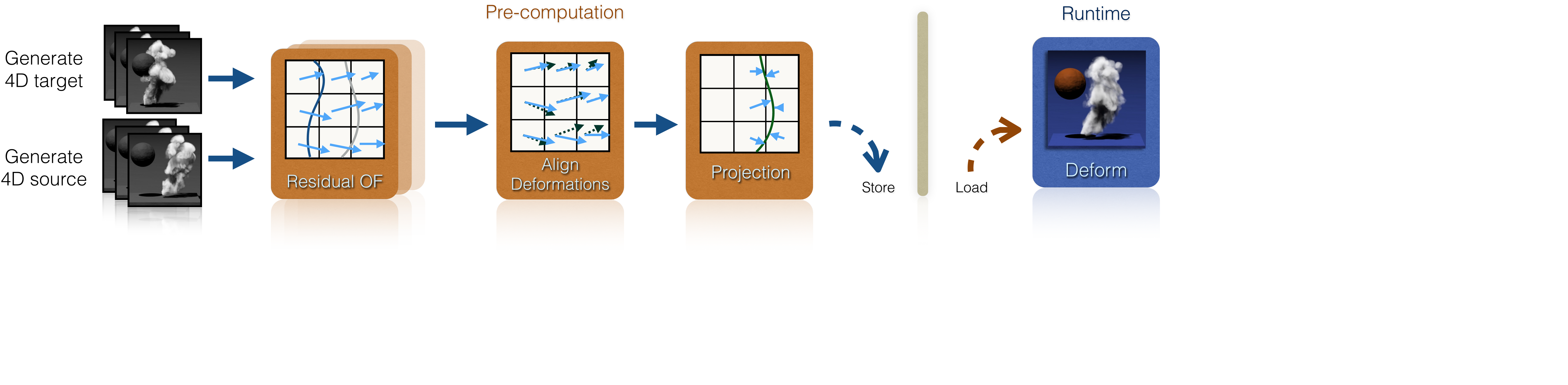}\end{center}  \vspaceImgMed
} \caption{ \label{fig:overview} 
	This figure shows an overview of our interpolation pipeline: 
	the input data is assembled into 4D signed-distance functions, which are analyzed with a
	5D optical flow (OF) solve. Our deformation alignment procedure
    combines the deformations from each pass into a single deformation field. 
	Our surface projection step then recovers small-scale
	details of the target surface, and the final deformations are stored.
	All that is necessary to generate a new simulation
	is to apply the pre-computed deformations (with a second alignment step).
} \end{figure*} 
Fluid simulations are established components of VFX production pipelines, 
with a variety of powerful solvers to choose from, including Eulerian methods \cite{Stam1999},
hybrids \cite{Zhu2005} and pure Lagrangian approaches \cite{muller2003particle}.
Surprisingly,
the tools for working with the large amounts of
simulation data produced by these solvers are extremely limited -- the simulation data is typically just
passed on to a rendering stage.
Any required change means re-starting a new simulation from scratch.

With this work, we target the reuse of simulation data
for the automated generation of in-betweens based on a {\em space-time deformation}. %
We precompute a matching of two 4D shapes.
Once it's computed, a user can freely choose any version in between the two extremes,
which can then be generated very efficiently, without starting a new simulation.

\revision{Beyond special effects, this is also highly interesting for 
interactive, data-driven simulations. Previous work in this area has demonstrated
the feasibility of precomputing state graphs that are suitable for games \cite{stanton2014srg}.
However, the graph quickly grows in size as all possible interactions have to be explicitly precomputed.
In such a setting, 
our method could be used to interpolate a few key variants, greatly reducing the state graph.
}

A key challenge for our work is to fully automate the matching process. 
In contrast to previous work on blending triangle meshes from liquids with user guidance
\cite{Raveendran:2014:blendingLiquids}, we target automatic matches of both smoke and liquid effects. The
only requirement for our approach is that a signed-distance function on a
regular grid is available. The Eulerian representation is especially useful for ensuring spatial and temporal
smoothness, which in practice translates into robustness.
\scndRev{This makes it possible to
directly work with the complex surfaces of splashy liquids, which
could otherwise cause misaligned deformations and incomplete matches.}

A central insight of this paper is the fact that signed-distance functions (SDFs),
which are readily available in many flow simulations, are a particularly
well-suited input for optical flow. 
With arbitrary data, the optical flow solve is strongly underdetermined, and will often yield suboptimal or unexpected motions.
However, the smooth gradients of SDFs give good results even for large deformations. 
Relying on the signed-distance property of the inputs also allows us to apply a
novel projection step to efficiently recover detailed correspondences.  In
combination with an iterated residual solving approach that we outline below,
our approach can find matches between significantly deforming flow surfaces
without any user input.
Additionally, the volumetric nature and robustness of our approach 
allow us to calculate matches for the complex shapes of 
swirling smoke clouds, or even completely different phenomena with changing
viscosities.

Specifically, our contributions are:
\scndRev{
%\vspace{-2mm}
%
\begin{itemize} \setlength{\itemsep}{1mm}
\item %
	\ a novel optical flow approach to register deforming space-time fluid surfaces given by SDFs;
\item \ a method to align multiple consecutive deformations into a single deformation field;
\item \ a multi-dimensional interpolation scheme for SDFs;
\item \ and an efficient projection to recover small details, in combination with residual iterations 
	and a robust volumetric error measurement.
\end{itemize} 
%\vspace{-2mm}
Together, these contributions lead to a practical algorithm 
for the robust registration and blending of complex volumetric phenomena.
}

\begin{figure*}[bt]
{\begin{center}
	\includegraphics[width=1.0\linewidth]{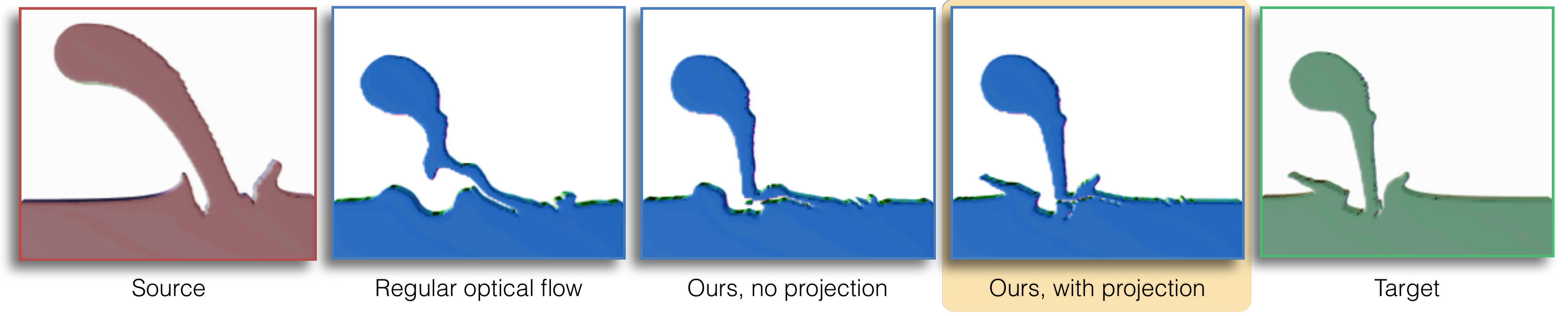}\end{center}\vspaceImgSm
} \caption{ \label{fig:ofSteps} 
	A 2D example illustrating the components of our pipeline.
	The images on the far left and right are the source and target inputs.
	The three center images from left to right are: the deformed surface
	after a single optical flow step, the result after the residual iterations, and 
	one including our projection step. While a single optical flow step does a mediocre job
	at registering the inputs, our final result closely matches the target surface.
} \end{figure*}

\section{Related Work}
\label{sec:relwork}

Fluid simulations in computer graphics were pioneered by the works introducing
stable and efficient grid-based solvers \cite{Foster:1996:RAL,Stam1999}, and have since
made significant progress. \fourthRev{The book by R. Bridson \shortcite{Bridson2016}} gives an excellent overview.  
We restrict
our discussion to Eulerian methods in the following, but %
powerful Lagrangian techniques are available \cite{IhmsenOSKT14a}. 
Our method could work with inputs from arbitrary solvers, as long as they
can be converted into signed-distance functions. For liquids, we use inputs
from the {\em Fluid Implicit Particle} (FLIP) approach \cite{Zhu2005}, which combines
particles and grid data. %
We also heavily make use of a semi-Lagrangian method \cite{Stam1999}, for which 
numerous extensions and improvements have been proposed, e.g., to increase accuracy
with a correction step \cite{Selle:2008:USM}, or to add small-scale turbulent detail \cite{Kim:2008:wlt}.

The aim of our method is closely related to fluid {\em guiding} approaches,
whose goal it is to influence the outcome of a simulation with respect to external,
and often non-physical goals.
While early works in this area have mostly focused on guiding
shapes \cite{ShiTamingLiquids,Thuerey:2006:SCA06dpfc}, recent techniques have introduced more subtle
techniques \cite{Pan:2013,Nielsen:2013:SWA} that are highly relevant for
practical applications. Overall, the aim of these methods differs from our
approach: they typically take a single goal surface as input, and then refine
or modify the result of a new simulation w.r.t. this input.  Our goal, on the
other hand, is to directly interpolate two or more inputs without running  new
simulations.

\revision{
Our method can also be seen as a way to precompute special reduced bases for flows.
In contrast to previous work in this area \cite{Treuille:2006:MRF,Kim:2013:SFR}, our aim is not to capture significant
flow motions to build a velocity basis, but to precompute correspondences between
the visible shapes of simulations. Thus, our method requires surfaces instead of motions as input. 
The method of Stanton et al.~\shortcite{stanton2014srg} is closer to our approach, and captures
complex liquid flows by precomputing a state graph that is adapted to player behavior. However, their algorithm
does not perform any interpolation on the precomputed data. As such, our contributions are orthogonal,
and could help to further reduce the state space in interactive settings.}\thirdRev{
The work by Ladicky et al. \shortcite{ladicky2015data} shares our goal to 
perform data-driven fluid simulations.
While we focus on interpolations of simulation data, they represent a broad class of particle interactions for 
Smoothed-Particle Hydrodynamics simulations with a regression forest. }

\revision{
The work of Stam and Schmidt \shortcite{stamISvel} explores different possibilities to calculate the surface velocity 
of a series of implicit surfaces. While we share the goal to reconstruct motions from implicit representations, 
our approach matches two or more space-time surfaces with
dense 4D deformations, instead of reconstructing normal and tangential 
surface motions of a series of 3D inputs. }

Our algorithm employs optical flow, which is a widely used approach
to retrieve motions from image data, and we will review several works
that are most relevant to our approach from this large field of research here.
The seminal work of Horn and Schunck \shortcite{horn1981determining}
has been investigated and extended in numerous ways. For our approach 
we use a hierarchical solve \cite{meinhardt2013horn}, and 
employ established best practices from the area \cite{ofsecrets2014}.
A good general overview can be found in books such
as the one by Wedel et al. \shortcite{wedel:2011:sceneFlow}.
Early on, Vedula et al. already proposed techniques to reconstruct 
dense 3D flows \shortcite{vedula1999three}, in their case based on multiple video streams.
Paragios et al. \shortcite{paragios2003non} used optical flow for non-rigid
registrations, to estimate a global rigid motion in combination with a local
deformation from image data.
Optical flow has also very recently been used in combination with fluid simulations
to reconstruct velocities from tomographic density data of real flows
\cite{Gregson:2014:divFreeof}.

While 2D and 3D variants are common, higher dimensional optical flow solves
are rare. One area where 4D solves have been successfully used
is registration of data, such as CT scans, in both space and time \cite{ehrhardt2007optical}.
In computer graphics similar techniques have been proposed to 
reconstruct captured performances with volume conservation \cite{Sharf:2008:SSR},
and to perform reconstructions in the presence of changing topologies and
inconsistencies \cite{popa2010globally}.  
Space-time data has also been useful for capturing complex phenomena such as trees \cite{Li:2013:AGPtrees}
and hair \cite{Xu:2014:DHChair}.
However, to the best of our knowledge, we are the first to use optical flow to
match multiple space-time data sets, thus effectively performing a 5D solve. 
All of the preceding methods typically perform a space-time
optimization on a single set of 4D data.

The work most similar to ours in spirit is the one by Raveendran et al.
\shortcite{Raveendran:2014:blendingLiquids}. 
They noticed that it is crucial to take into account both space and time,
to give the optimization enough freedom to find a good match.
While they share our goal
of computing a space-time registration of surfaces, there are several
important differences. First, we target Eulerian SDF surfaces, which
are more widely used for surface tracking than meshes. Second, our grid-based
representation allows for very efficient regularization, and thus more robust
matching. Also, no helper data structures are needed for closest point lookups (we
can simply follow the gradient of the SDF), the implicit representation
inherently handles topology changes, and many operations such as
intersections with time planes are trivial and very efficient in our setting.
\scndRev{
Note that the mesh-based approach \cite{Raveendran:2014:blendingLiquids}
could be run on triangulations of SDFs. However, in this case no vertex correspondences
exist, leading to a degradation of quality.}
Lastly, while a central component of the mesh-based approach was user guiding, 
we did not require any additional information from a user. Thus, we can run
our algorithm automatically on a large number of inputs.  

\section{Method}
\label{sec:method}

Given a sequence of surfaces over time from a simulation run, we
concatenate these surfaces as time slices of a 4D volume. 
For a liquid simulation, we simply use the surface of the liquid phase, and for a smoke
simulation we use an iso-level of the density volume (details will be given in \myrefsec{sec:results}).
As the 3D signed-distance values do not contain any information
about proximity in time, we calculate 4D signed-distance values for this
surface. 
We use an equidistant discretization for the spatial and temporal
dimensions of our 4D data, and all dimensions are weighted equally when
computing the magnitudes of 4D vectors\footnote{
$|(1,0,0,0)^T| = |(0,0,0,1)^T|$.
Thus, at 
$\mathbf{p}=(x,y,z,t)^T$, cell neighbors in 
space and time have equal distance: 
$|\mathbf{p}-(1,0,0,0)^T|=|\mathbf{p}-(0,0,0,1)^T|$. \fourthRev{We consider inviscid flows; when solving
with viscosity, this could be used to relate space and time across simulations.} } .
Our algorithm takes {\em two} of these 4D volumes from different simulation
setups as input: $\imVc_1$ and $\imVc_2$. It computes a dense field of
four-component deformation vectors $\velV$, which maps points of $\imVc_1$ onto $\imVc_2$.
This deformation has a clear direction. In the following we will assume
this direction to be from  $\imVc_1$ to $\imVc_2$, but it is possible to
compute the inverse deformation in a separate step.

In the following, we first review the most important concepts of optical flow,
before introducing our extensions.

\subsection{Optical flow}

The goal of the optical flow step is to compute
the deformation $\velV$ to transform one input into the other one. 
\fourthRev{The whole non-linear, and potentially long range, deformation
is de-composed into several smaller, linearized steps, the first few of which we solve with optical flow.}
For brevity, we will only give a brief overview of
the derivation, and then focus on the discrete version of the optical flow solve.

Notation: For a continuous value (e.g., $\imVc$)
we denote its discrete counterpart with a bold symbol (such as $\imV$).
For matrices we will use bold, upper-case letters (e.g., $\Mat{A}$). 
We will use one or more $'$ to denote intermediate results. Thus, $'$ will never indicate a derivative.

We make use of a hierarchical variant of the commonly used
Horn-Schunck~\shortcite{horn1981determining} algorithm.  The algorithm is
motivated by the so-called {\em brightness-constancy assumption}.  I.e., values
in an image (or voxels in a volume) move around, but do not change in
magnitude.  Usually, optical flow considers two images taken at different
times. \fourthRev{We instead use optical flow to compute correspondence 
between the space-time surfaces of different simulations.
We use a parameter $r$ to reflect the change of the simulation inputs.}
$r$ could potentially be any parameter changing
the outcome of a simulation, e.g., the initial position of a drop, viscosity,
or even cell size.
\thirdRev{
Given an input $\imVc(x(r),r)$ that moves in space-time w.r.t. $r$,
the brightness-constancy assumption can be expressed as
$\text{d} \imVc / \text{d} r = 0$.
Thus, we require $\imVc$ to be constant as we change $r$.
Applying the chain rule yields
\begin{eqnarray} \label{eq:dataTerm}
 \frac{\partial \imVc}{\partial r} + \frac{\partial \imVc}{\partial x} \cdot   \frac{\text{d} x}{\text{d} r}  = 0 \ .
\end{eqnarray}
This is directly in line with the commonly used material derivative for
advection in fluid solvers.  The main difference here is that we want to
recover the motion of $\imVc$, while fluid solvers typically compute this motion, 
and aim for computing its effect on $\imVc$.
In our setting, the change of position $\frac{\text{d} x}{\text{d} r}$ corresponds to the deformation $\velVc$.
For further details of the derivation, we refer
interested readers to Section~4 and Appendix~A of the original Horn and Schunck paper \shortcite{horn1981determining},
or books on optical flow, such as the one by Wedel et al.~\shortcite{wedel:2011:sceneFlow}.

The second term of \myrefeq{eq:dataTerm} represents a non-linearity that
turns out to be difficult to linearize. Typical fluid solvers take care to compute 
these terms as accurately as possible with specialized algorithms \cite{Selle:2008:USM}, and the non-linearity
is similarly challenging for inverse problems such as optical flow. 
Below, we will discuss
several steps to robustly retrieve solutions. 
}

\scndRev{
As solving for brightness-constancy alone is strongly underdetermined, a variety of regularizers
have been proposed. We employ the two most common regularizers:
one penalizing non-smooth solutions, and a second one to favor
deformation vectors with small magnitudes (a so-called {\em Tikhonov} regularizer).}
The resulting problem is now formulated as an energy minimization:
the goal is to compute a deformation field $\velV$ which
minimizes the data term \myrefeq{eq:dataTerm}, and the two
regularizers. 
\fourthRev{
We calculate the deformation by minimizing a weighted sum 
of the energy terms
\begin{eqnarray} \label{eq:ofContA}
\underset{\velVc}{\mathbf{min\;}} E_{\text{data}}(\velVc) + \beta_S E_{\text{smooth}}(\velVc) + \beta_T E_{\text{Tikhonov}}(\velVc) \ .
\end{eqnarray}}
\fourthRev{
The discrete version of the energy to be minimized
in a least-squares sense is given by
\begin{eqnarray} \label{eq:ofCont}
E_{of}(\velVc) = \frac{1}{2} \! \int_\Omega \! \left| \imVc_r + \velVc \cdot
\nabla \imVc \right|^2  
\!+ \beta_S \! \sum_j |\nabla u_j|^2 
\!+ \beta_T |\velVc|^2
d\Omega \ 
\end{eqnarray}}
where  $\beta_S$ and $\beta_T$
represent scaling factors to control the relative importance of 
the three terms, and $\sum_j$ sums over the four dimensions. 
While $\imVc_r$ usually denotes the time derivative for regular optical flow applications,
here it corresponds to a {\em data derivative}: 
the change induced by the aforementioned parameter $r$, which could be any
external control knob modifying the simulation (e.g., a moved inflow position).
Thus, our formulation effectively considers five dimensions.
\fourthRev{We compute the spatial gradient $\nabla \imVc$ with $\imVc_2$.}

The discrete minimization of \myrefeq{eq:ofCont} yields a system of linear equations 
\begin{eqnarray} \label{eq:ofmain}
\matof ~ \velV = \Vect{b} \  ,
\end{eqnarray} 
which we solve for $\velV$.
The matrix $\matof$ contains the discretized energy terms, and
is given by:
\begin{eqnarray}
  \matof = \GradPhiA^T \GradPhiA + \beta_S \sum\nolimits_j  \Mat{L}_j + \beta_T \Mat{I} 
\end{eqnarray}
The terms, from left to right, correspond to the aforementioned
data term, smoothness, and Tikhonov terms. 
$\GradPhiA$
denotes an $n\times 4n$ matrix containing the discrete gradient of $\imV_2$,
and $\Mat{L}$ denotes the discrete Laplacian.
 The right hand side of the linear system is
$\Vect{b} = -\GradPhiA^T \imV_r $,
where a finite difference with a normalized step size is used to compute $\imV_r = \imV_2 - \imV_1$.
The gradients in $\matof$ and $\Vect{b}$ are 4D vectors along spatial as well as time dimensions.

\thirdRev{
An inherent difficulty here is to linearize the non-linear terms of \myrefeq{eq:dataTerm}.
As is common practive for optical flow methods, we use first-order approximations
for the corresponding derivatives \cite{wedel:2011:sceneFlow}.
As a consequence, this basic optical flow solve works well for small
deformations (on the order of 1-2 grid cells), but fails to recover larger
motions.
We will later on propose an iterative scheme, which we combine with
the commonly used hierarchical procedure to recursively re-sample and deform
the data on coarser resolutions \cite{meinhardt2013horn}. Both methods
help to reduce the inaccuracies of the first-order approximations.
The hierarchical scheme does this by solving the equations on different spatial
scales, while our iterations combine the results of multiple hierarchical solves.} 

A central insight of our work is that using SDFs as input for the optical
flow solve is paramount for retrieving high-quality deformations.
For smoke volumes it might seem natural to calculate optical flow
directly on the density data, as we typically  perceive smoke clouds 
to be highly textured. Likewise, one might
try to use fill-fraction values for liquid simulations. 
However, both cases will yield clearly suboptimal results.
This problem is the well known {\em aperture problem}, %
which arises from the brightness-constancy assumption:
in regions with uniform intensities the motion is completely ambiguous - any
point could move to any other one. 
A typical simulated smoke cloud has zero values outside, and most likely saturated
values inside. These regions contain zero information about the motion of the
underlying fluid.
In typical optical flow applications,
this problem is alleviated with regularization, however, it is preferable
to minimize the ambiguity in the first place. This is where SDFs are highly 
beneficial: by definition, they have values that change with distance to the 
implicitly defined surface at the zero level set.
\thirdRev{ While there is still ambiguity w.r.t. to the tangential direction of the surface,
a band of values normal to the surface can now be robustly matched.}

\begin{figure}[bt]
{\begin{center}
	\includegraphics[width=1.02\linewidth]{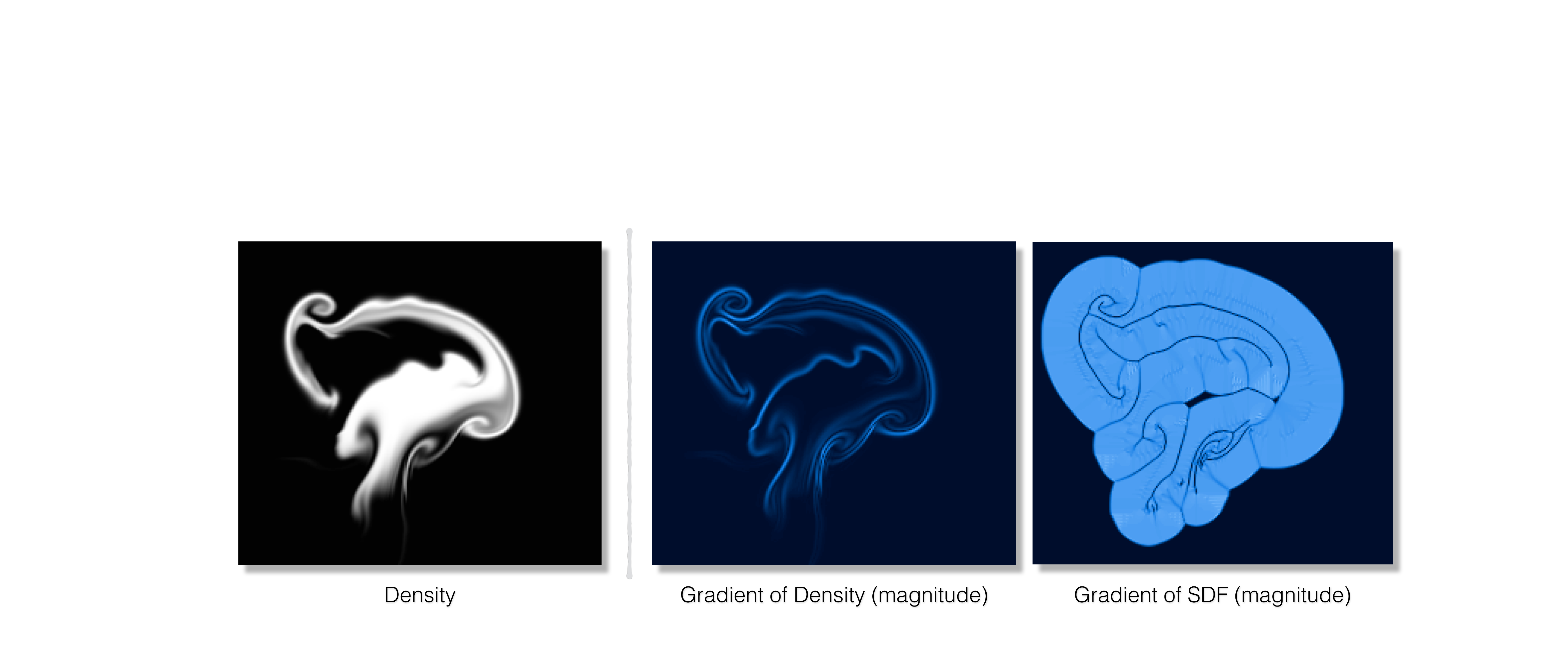}\end{center}\vspaceImgSm
} \caption{ \label{fig:sdfVsSmoke} 
	\revision{
	These images highlight the importance of using an SDF for the optical flow solve. 
	An example input density is shown on the left, while the next two images show magnitudes of
	gradients in blue. 
	\scndRev{Larger gradients are shown with brighter shades of blue. The center image
	was generated from gradients calculated based on the input density, while the right image
	was calculated from an SDF of the 0.5 level set of the input. 
	As these gradients make up the block-diagonal band
	of the optical flow matrix $\mathbf{A}_{of}$, reliable gradients are crucial,
	and the area of bright blue values in the middle and right images directly 
	indicates how many reliable gradients can be used to assemble the matrix in each case.}
	}
} \end{figure}

In the optical flow solve, the motion is expressed in terms of the gradient of
the input data.  Sub-optimal input data, with saturated or empty regions, has
insufficient, and ambiguous gradient information as a consequence, and spatial
gradients exist only in a relatively small band. 
The sparsity of such data is illustrated for a smoke cloud 
in \myreffig{fig:sdfVsSmoke}. The center image shows gradient magnitudes, and the predominantly
black areas of the image indicate a complete lack of gradient information.
In contrast, an SDF has smooth and reliable gradients even far away from the
surface (effectively, as far as distance information was generated).
\myreffig{fig:sdfVsSmoke} (right) shows gradient magnitudes for an SDF
generated for the $0.5$ level set of the smoke cloud on the left. 
The uniform
bright blue color indicates gradients of length one. The original surface is not shown. It lies
right in the middle of the thick blue band. The gradients could easily fill the whole
image. The drop off to zero only happens because the SDF was truncated at a distance of 15 cells 
in this example. The medial axis of the SDF is also visible as black lines in the interior 
of the shape.

The gradients of the input data make up the core diagonal blocks of matrix 
$A_{of}$ of \myrefeq{eq:ofmain}. As such, they are crucial for a well-posed optical flow problem.
The matrices from SDF data typically have much more reliable deformations,
as the gradient terms on the diagonal guide the solution in a broad
band around the surface. 
As additional benefit, using 4D SDFs allows us to match small and fast moving 
structures: even fast moving parts of the surface will have a large {\em halo} of 4D distance values, 
on which the optical flow can operate. 

\subsection{Residual Optical Flow}
\label{sec:multistage}

\scndRev{
If a first deformation $\velV^1$ calculated by optical flow brought us closer to the target,
we can apply the current deformation, and perform another solve for the remaining difference, i.e. the residual,
yielding a second deformation $\velV^2$.
This works particularly well in a setting where we have a clear target without noise
or other measurement errors. 
Here this goal state is exactly the target SDF $\imV_2$.
We can continue to refine the deformation with 
residual iterations until either a
computational budget is exhausted, or the refinement does not yield gains in quality anymore.
To combine the sequence of deformation fields $\velV^{1..n}$ from each solve into a single one 
we employ the {\em deformation alignment} from \myrefsec{sec:alignment}.
The difficulties that arise when combining multiple Eulerian deformations,
as well as our solution will be explained in full detail there.
As a result, we retrieve a single deformation that aligns the two surfaces significantly better
than only one optical flow solve (see \myreffig{fig:ofSteps}, middle).
}

Conceptually, these residual iterations are important, as the optical flow
solve performs the aforementioned linearization of \myrefeq{eq:dataTerm} around the input states.
The iterations re-linearize the problem closer and closer to the target state,
and in combination with the hierarchical solve can significantly improve the
final quality of the deformation.
\thirdRev{
Previous works have considered directly solving the non-linear problem with
Newton's method \cite{zikic2010}, but generic non-linear solvers
are typically outperformed by methods that employ specialized algorithms for advection
(e.g., in combination with the hierarchical scheme that we employ \cite{meinhardt2013horn}).
Other variants of optical flow solvers
re-formulate the equations by splitting of the deformation into a fixed part
and an incremental deformation.
The increment is retrieved by repeated optical flow solves \cite{Zach07aduality},
typically also in combination with a pre-warping step that uses an advection algorithm.

In contrast, our approach is based on aligning the different deformation fields
with the method from \myrefsec{sec:alignment}. 
While our method yields results that are similar to the incremental variants, 
our method has the advantage that additional
deformation fields could be incorporated by alignment. 
E.g. future extensions of our algorithm could improve the matching
of two inputs by post-processing the deformation computed by optical flow. 
Our alignment could then be used to combine the optical
flow deformation, and one or more post-processing deformations into a single one.
An additional smaller advantage is that given a function for aligning deformations,
the residual iterations can be easily implemented. Thus, this approach seamlessly integrates
into our interpolation framework.}

\subsection{Surface Projection}
\label{sec:surfaceproj}

The optical flow step does a good job at robustly detecting smooth
large scale motions, but its inherent regularization prevents it from matching
fine details at the surface. To retrieve this detail, we can 
leverage knowledge about the input data: it is a signed-distance function, and we know that the
source surface should ideally deform to end up exactly on the target surface. If
the surfaces are sufficiently close, such a deformation can be easily obtained 
by marching along the negative gradient of the SDF. 
This projection is inspired by techniques from mesh-based registration
\cite{Bojsen-Hansen:2012}. We will show that it can be adapted to a volumetric
setting, and leads to considerable gains in quality regarding small-scale
detail.

Given an SDF deformed by a deformation
from the optical flow step, we perform a bisection search for each cell
along the gradient of the target SDF, until we find the spot
on the target where it matches the iso-surface value of the source cell.
This line search has the advantage of being trivially parallel, as the result depends purely
on the deformed SDF and target SDF.
We consider the projection as an update step for a current deformation 
estimate $\velV^{k}$:
\begin{eqnarray} \label{eq:projectCell}
\velV^{k+1}(\mathbf{x})  = \velV^{k}(\mathbf{x}) + 
  \text{bisect}\big( \mathbf{x} + \velV^{k}(\mathbf{x}) , ~ -\mathbf{n}(\mathbf{x} + \velV^{k}(\mathbf{x})  ) ~ \big) \ ,
\end{eqnarray}
where the function $bisect$ performs the actual search from the start position (first parameter) along the direction passed as the second
parameter. $\mathbf{n}$ in this case denotes the normalized gradient of the target $\imV_2$
and is inverted for points inside the volume.

\begin{figure}[bt!]
	\includegraphics[width=0.480\textwidth]{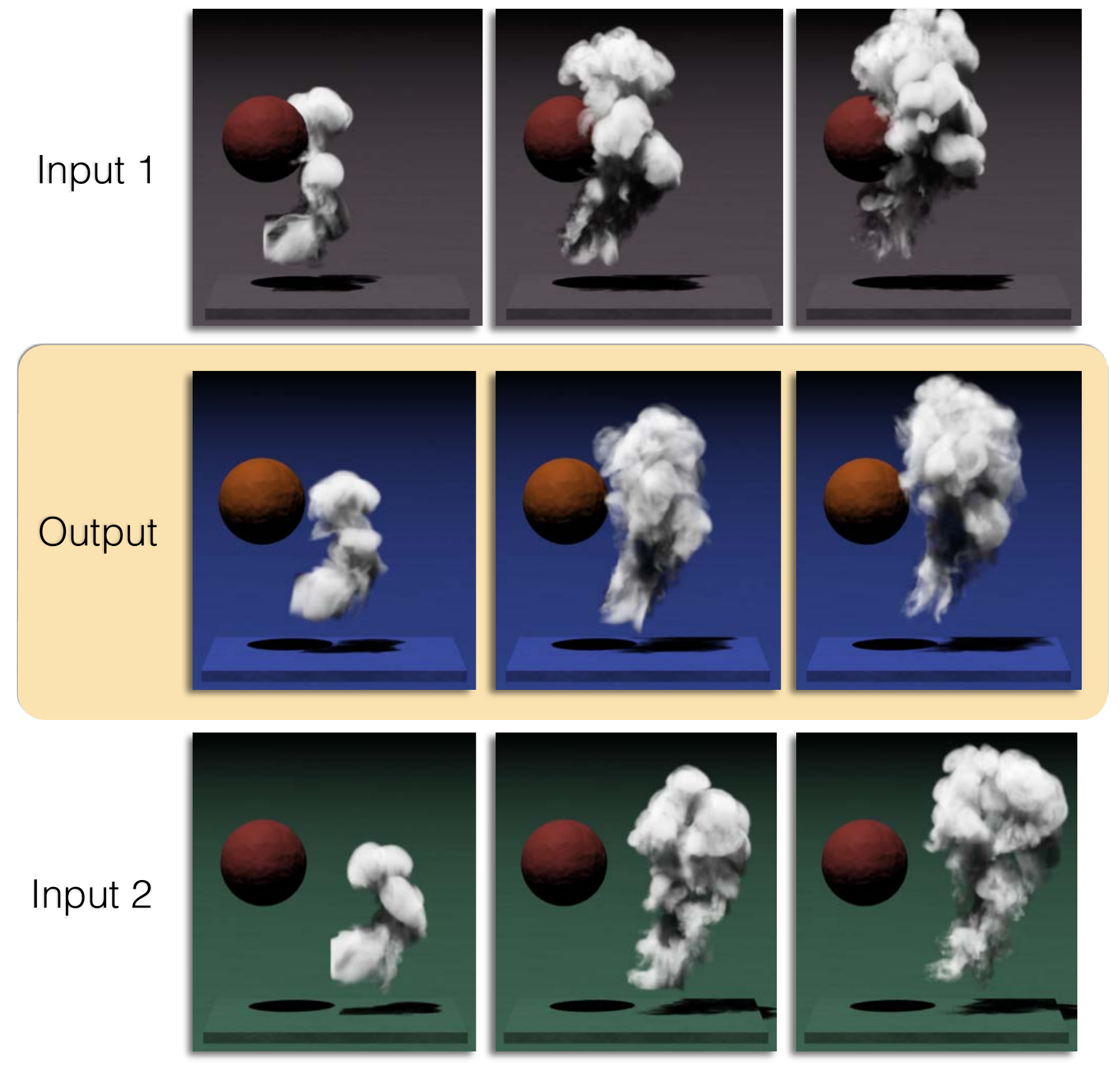} \vspaceImgMed
   \caption{ \label{fig:smokeComp}
	An example deformation for a smoke simulation. The inputs on the top and bottom are matched
	with {\em FlOF} and interpolated for a weight of $0.5$ in the middle.  
	Our interpolation recovers the translation along $x$, as well as the deformation of the cloud. }
\end{figure}

The completely independent calculation of the projection
for each cell can yield different deformations in regions of quickly changing target normals, and
the calculation of the direction is unreliable on the medial axis of the SDF.
However, we found that the projection step can yield very good results if it is restricted
to a narrow band of width $\dproj$ around the surface, and its output is smoothed %
using a Gaussian of size $\sigma_{\text{proj}}$.
To propagate the effect of the projection into the volume around the surface,
we extrapolate the deformation values before applying the smoothing.
We do this with $\dproj$ explicit iterations \cite{JeongW08},
fading out the deformation over the course of the iterations.
As illustrated in \myreffig{fig:ofSteps} the combined deformation snaps the
input very tightly to the target surface, recovering small-scale details in this way.

Note that this projection step solves essentially the same problem as the optical flow step.
However, due to the aforementioned difficulties, it is no replacement for it. Instead,
the projection only gives good results once a suitable overall match has been obtained by 
optical flow. %

\begin{algorithm}[bt!]
\footnotesize
 {\bf function} FlOF($\imV_1, \imV_2, \velV$) { \\
 	\If{ $\text{gridsize}(\imV_1)$ $\ge s_{\text{max}}$ }{
 		$\imV_1', \imV_2', \velV'$ = Downsample($\imV_1, \imV_2, \velV$) \\
 		$\velV'$ = FlOF($\imV_1', \imV_2', \velV'$) \text{// recurse}  \\
 		$\velV$ = $\velV$ + Upsample($\velV'$)
 	}
  	\For{$l=1$ to $l_{max}$}{
 		$\imV_1''$ = advect($\imV_1, \velV, 1$) \\
 
 		$\velV_l$ = opticalFlow( $\imV_1'', \imV_2, \sigma_{\text{of}}$ ) \\
 
 		$\velV_{\text{tmp}}$ = $\velV$ + alignVelocity($\velV_l$) \text{ // see Sec.~\ref{sec:alignment} } \\
 
 		\If{e(\text{advect}($\imV_1,\velV_{\text{tmp}}$ , 1),$\imV_2$) $\le$ e(\text{advect}($\imV_1,\velV$ , 1),$\imV_2$)}{
 			$\velV = \velV_{\text{tmp}}$ \\
 			$\sigma_{\text{of}} = 3/4 \, \sigma_{\text{of}}$ 
 		} \Else {
 			break \\
 		}
 	}
 
 	\If{ is finest grid$(\imV_1)$ }{
 		\For{$k=1 \text{ to } k_{max}$}{
 			$\velV$ = $\velV$ + project( $\imV_1, \imV_2, \velV, \sigma_{\text{proj}}$ ) \\
 			$\sigma_{\text{proj}} = 3/4 \, \sigma_{\text{proj}}$
 		}
 	}
 	\KwRet{ \velV}
 }
 
	\vspace{5pt}
   \caption{ \label{alg:mainAlg}
   	Pseudo-code for our full algorithm. Parameters are given in \myreftab{tab:params}.
    }

\end{algorithm}

\begin{figure}[tb!] { 
	\begin{center} \includegraphics[width=1.00\linewidth]{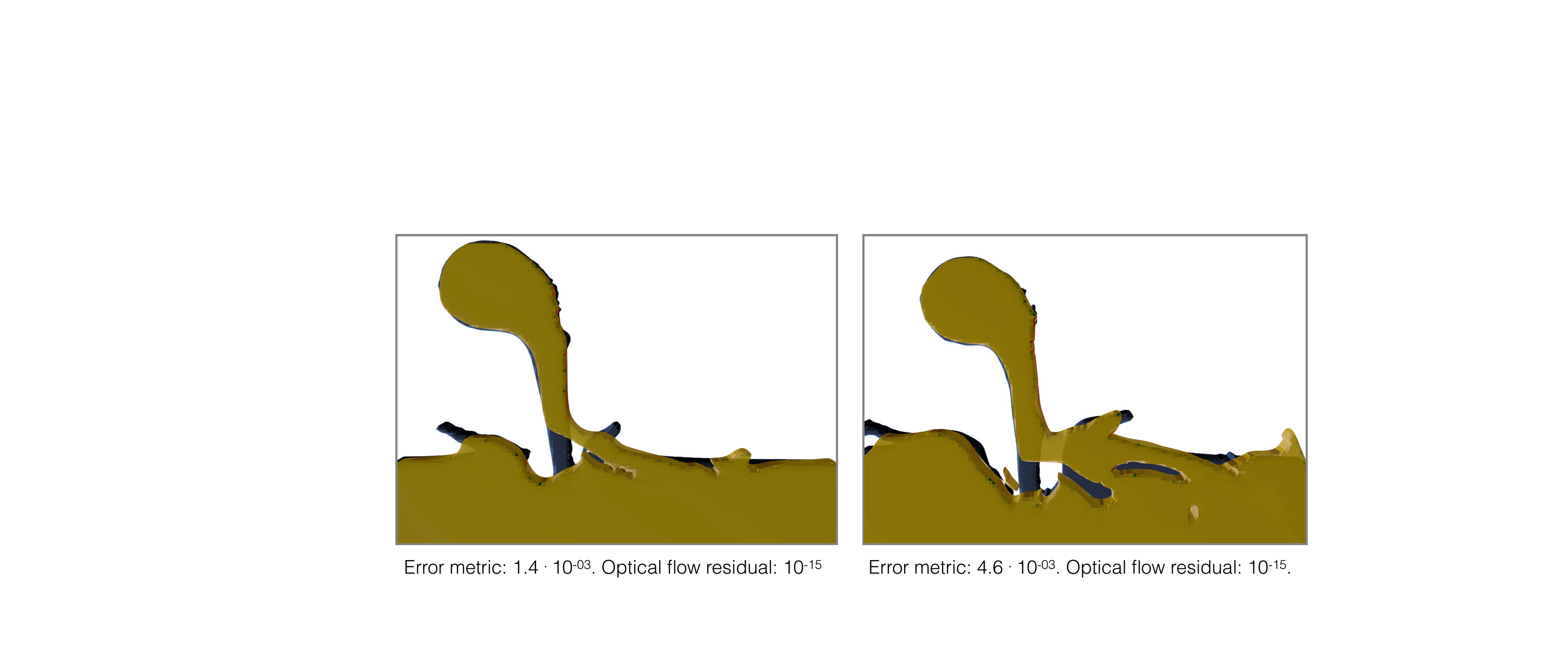} \end{center} \vspaceImgMed
} 
\caption{ \label{fig:errorMetric} \thirdRev{This image illustrates the advantages of our error metric
	for two different time steps of a two-dimensional input: while
	both final deformed surfaces (in yellow) have very similar optical flow residuals close to zero, our error 
	metric detects a significant difference. The higher error is caused by the broken up surface in the center region of the right image. 
	The target surface is shown in dark blue in the background.}}
\end{figure}

\subsection{Error metric}
\label{sec:error}

A last component that is important for combining the aforementioned
techniques is a reliable metric to quantitatively evaluate a deformation.
Unlike settings where optical flow is traditionally employed (e.g., to calculate motions
in video data), our synthetic data has no occlusions, or
motion blur, which motivates our choice of metric.
Alternatively, the energy from \myrefeq{eq:ofCont} could be used
to evaluate the final quality, but we found that discarding the regularization terms,
and putting emphasis on the volume itself yields results
that are more in line with the perceived alignment of the surfaces.
While an algorithm to directly minimize the following metric is imaginable,
we leave this for future work, and purely use the error metric 
as a stopping criterion for our optical flow and projection iterations.
We define the error metric as the integral of an indicator function $h$ over the domain. 
In a discrete setting given %
two input SDFs $\imV_1$ and $\imV_2$ (which we assume to be calculated 
for a normalized cell size of 1), this becomes: 
\begin{eqnarray} \label{eq:error}
 e(\imV_1,\imV_2) = \sum_{\Vect{x}} h( \imV_1(\Vect{x}) , \imV_2(\Vect{x}) ) V \ ,
\end{eqnarray}
where $V$ is the volume of a cell. 
The question which function to choose for $h$ has been explored in various settings, e.g., 
for mesh similarity \cite{Cignoni:1996:MME}, or within optimizations
for shape registration \cite{cremers03shapedist}.
We have found that a metric detecting non-matching volumes is especially
important in our setting: 
\vspace{-0.2cm}
\begin{eqnarray} \label{eq:errIndi}
h(s_1,s_2) = 
\begin{cases}
0                          & \text{ if } \text{sgn}(s_1) = \text{sgn}(s_2)\\
\text{min}(1, |s_1 - s_2| / \Delta x ) & \text{ otherwise. } 
\end{cases}
\end{eqnarray}
\vspace{-0.2cm}

This metric has a value of at most one where the two SDFs disagree, and is zero where
the SDFs agree, but still detects sub-cell shifts. 

We have tried alternate metrics such as a simple difference of the two SDFs
($h=|\imV_1-\imV_2|$), the {\em metro} distance mentioned above
\cite{Cignoni:1996:MME}, or the energy of the optical flow solve, but we found that our version from
\myrefeq{eq:errIndi} is more reliable in practice. 
\fourthRev{
What turned out to be most important is a metric
that puts more emphasis on the surface region.
When using error metrics based on direct differences of the two inputs, large distance
values easily led to undesirably large error values.} Thus, 
the other metrics tend to introduce large errors when small pieces of
the surfaces disagree (e.g. drops far from the bulk volume). However, this is
visually not as crucial as a good match of the large scale volumes of the
liquid surface, which is why our metric puts more emphasis on the latter.

\thirdRev{The comparison in \myreffig{fig:errorMetric} demonstrates that our metric
is more reliable for detecting differences than the optical flow energy itself.
In both cases, a deformed surface is shown in yellow, while the surface
of the target is shown in dark blue in the background. The optical flow solves
were performed with high accuracy, and have a very small residual of $10^{-15}$ for both cases. 
Visually, a difference is clearly noticeable, but the averaged optical flow energy
cannot detect the remaining differences. Instead, our error metric
detects a residual difference for both versions, and
yields a three times higher value for the right input. To purely illustrate the effect
of our error metric versus the optical flow energy, we have disabled the surface projection
and residual iterations for this example.}

\begin{figure}[b!] { 
	\begin{center} \includegraphics[width=1.00\linewidth]{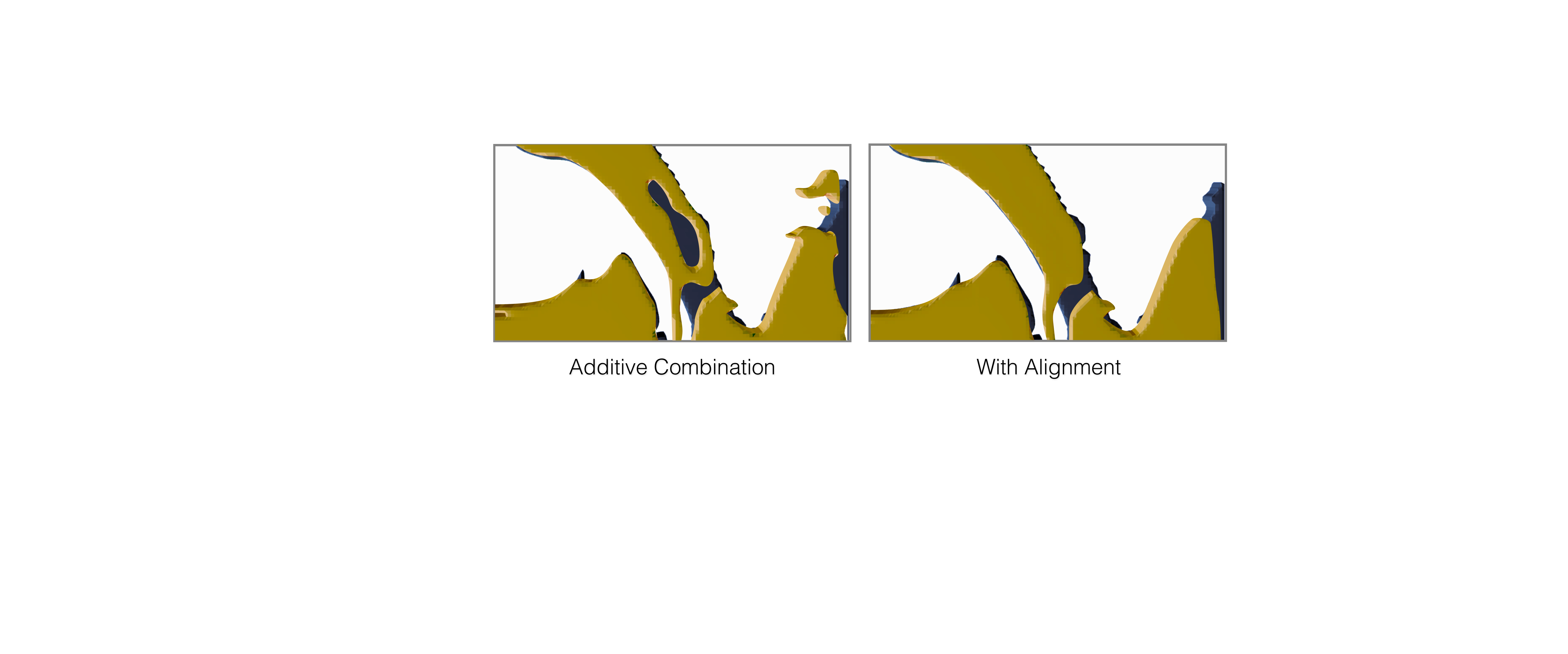} \end{center} \vspaceImgMed
} \vspace{-0.21cm}
\caption{ \label{fig:alignmentExample}  The effects of our deformation alignment for a practical example:
	Using an additive combination (left image), structures can break up
	or artifacts can develop, e.g. in the top-right corner of the
	left image. The target surface is shown in dark blue in the background.}
\end{figure}

\subsection{Full Algorithm}

\myrefalg{alg:mainAlg} outlines the integration
of our residual OF and projection steps into the hierarchical
optical flow scheme. 
Here the solution of the optical flow solve is smoothed with a Gaussian kernel of size $\sigma_{\text{of}}$, before
passing the deformation on to the next higher level (or the caller).

For the final algorithm we adopted another
strategy from the mesh-based registration area: 
it is more robust to first match larger scales, before moving to smaller ones. We can
do this within a single level of the hierarchy by varying the blur kernel radius.
Thus, when iterating either the optical flow or projection, we start
with a relatively large kernel, and then reduce its size by a factor of $3/4$
after each step. We use our error metric from \myrefsec{sec:error}
to check whether the last optical flow solve leads to an overall improvement,
or whether it is preferable to stop iterating.

\scndRev{
The full matching algorithm, which we denote with {\em FlOF} (for FLuid Optical Flow),} is summarized in \myrefalg{alg:mainAlg}.  
Initially, our algorithm is called with the two input SDFs,
and a zero velocity $\velV=0$. The 
{\em advect}$(\bf{a},\bf{v},\alpha)$
function applies the deformation $\bf{v}$ to $\bf{a}$ with weight $\alpha \in [0..1]$.

\section{Deformation Alignment}
\label{sec:alignment}

\scndRev{
In the following we will describe our new approach to align multiple consecutive
Eulerian deformations so that they can be merged into a single deformation field. \fourthRev{This alignment
is useful for the residual iterations above, each of which produces a separate deformation.
It is also crucial for the higher-dimensional interpolations of \myrefsec{sec:blend},
to align the results of multiple {\em FlOF} solves.}
In both cases it is highly preferable to compute a single, smooth linear
deformation from source to target, instead of storing and applying sequences of
deformations. 
The importance of the alignment for a practical 2D deformation example can be seen in \myreffig{fig:alignmentExample}.
We will first describe the alignment for two deformations, 
before considering arbitrary sequences. }

\revision{
Given an input surface $\imVal$ and two deformations $\velAl_1,\velAl_2$ with
scaling factors $\alpha_1,\alpha_2$, respectively, we assume
for now that each deformation is applied with a semi-Lagrangian lookup
$\imVal'(\posAl) = \imVal(\posAl - \velAl(\posAl) )$. 
Our approach would likewise work with higher order methods \cite{Selle:2008:USM}.
Here the $'$ indicates that $\imVal'$ is a temporary value, which is only required for the next
calculation step, and can be discarded afterwards. 
We arrive at the final configuration $\imVal_{\text{dst}}$ with
\begin{eqnarray} \label{eq:alignAdvect2}
 \imVal'(\posAl)   &=& \imVal(\posAl - \alpha_1 \velAl_1(\posAl) )  \nonumber  \\ 
 \imDstal(\posAl) &=& \imVal'(\posAl - \alpha_2 \velAl_2(\posAl) )
\end{eqnarray}
Our goal 
is to reach $\imDstal$ with a single advection
calculation using the unknown deformation $\veldst$:
$\imDstal(\posAl) = \imVal(\posAl - \veldst(\posAl) )$.
} %

\begin{figure}[bt!] {  \vspaceImgMed
	\begin{center} \includegraphics[width=0.99\linewidth]{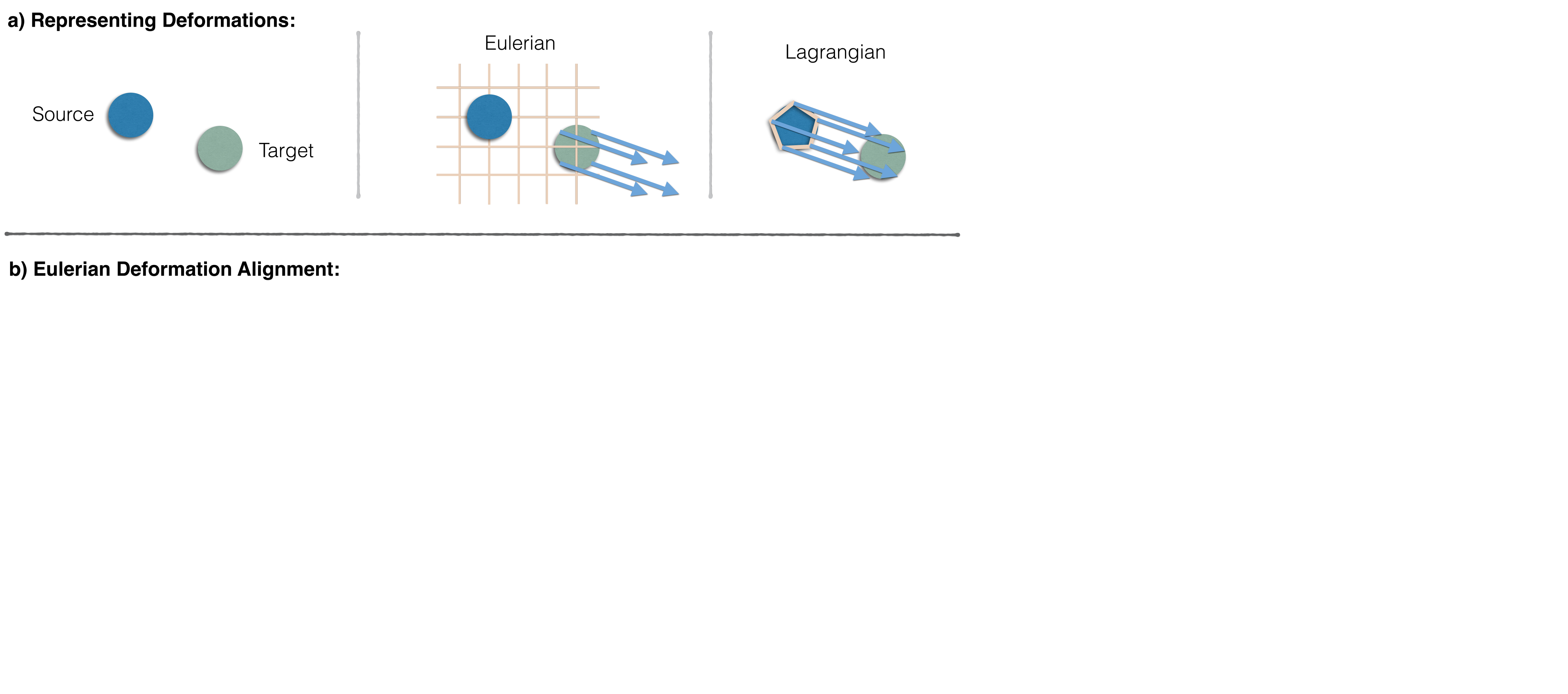} \\ \includegraphics[width=1.00\linewidth]{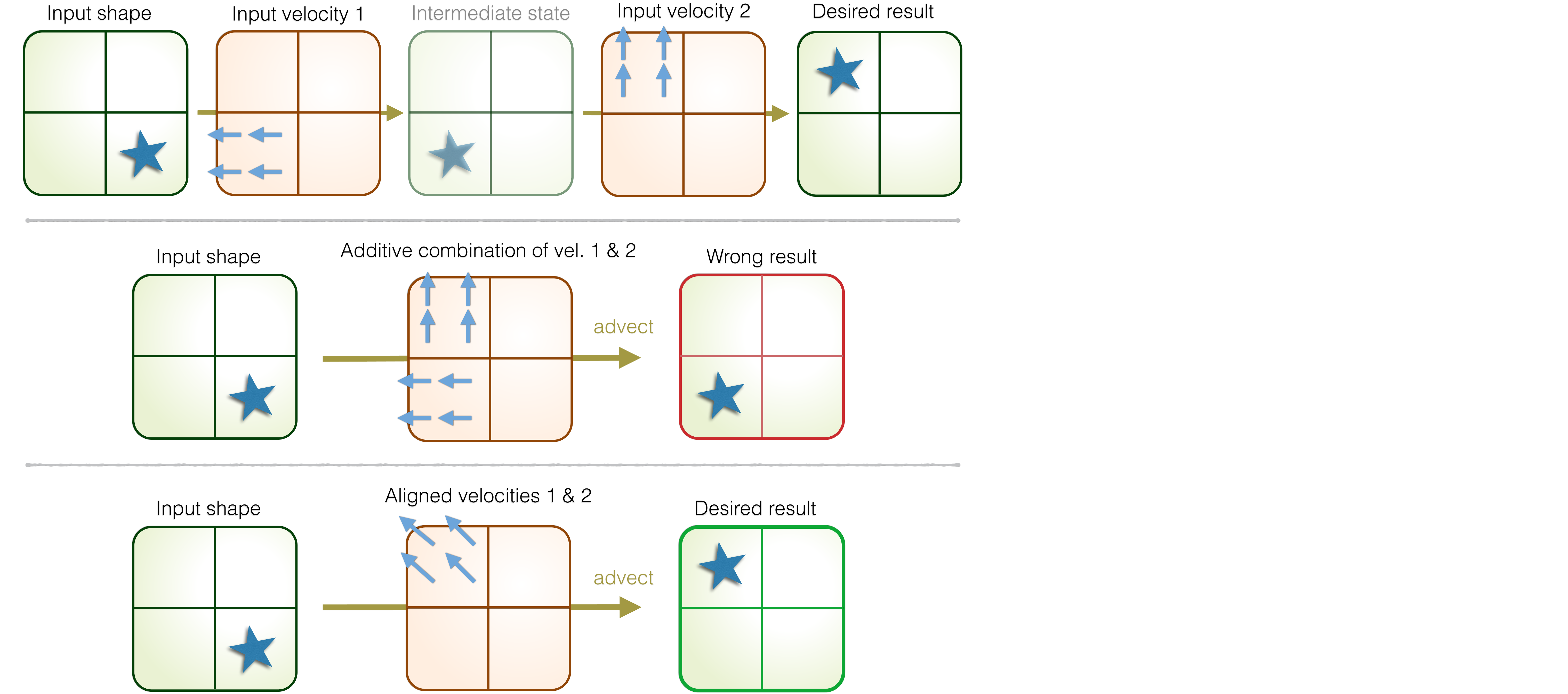} \end{center}\vspaceImgSm
} \caption{ \label{fig:defoCombine} 
	a) A sample deformation (left) can be represented in 
	an Eulerian way (middle, only relevant deformation vectors shown) or a Lagrangian way (right).
	For the latter, the deformation vectors are located at the source position (i.e., at vertices or particles), while
	the Eulerian setting we are using requires the deformation to be stored in the cells of the target position.  \\
	b) An example highlighting the importance of aligning Eulerian deformations:
	the top row shows the input SDF (a star shape), that is deformed with two consecutive deformations (orange fields).
	An additive combination of the two deformations effectively ignores the second one,
	while the result of our alignment (bottom) gives the desired result with a single step.
} \end{figure}

\thirdRev{
Combining these deformation is difficult primarily due to the
Eulerian representation. Eulerian advection methods are typically {\em backwards}
looking, i.e., pull data towards a sample location.}
Thus, if we consider a surface to be deformed, the deformation vectors
acting on the surface are not located at the actual surface positions, but instead in the
target region where the surface should be moved to. For every spatial position, e.g., every
cell in a grid, we perform a backwards lookup along the velocity direction to locate
which data should be moved to the cell under consideration.
This is in contrast to a Lagrangian representation, where the deformation is
stored directly at the surface, and each surface point is moved forward with
its deformation vector.
This difference is illustrated in \myreffig{fig:defoCombine}(a).
While combining two deformations is trivial in Lagrangian settings -- they
can simply be added -- this approach yields incorrect results for Eulerian representations.
Intuitively, the problem is that the deformations for a single surface point
are located at two different locations, one for each deformation.

It is also important here that our Eulerian deformations vary spatially.
Simple additions would work for cases such as uniform translations, but the 
deformations we are interested in often move different parts of the surface
to different target locations, and as such are typically far from uniform.
In the following, we will explain how to align two or more of these
deformations, such that they can be merged into a single Eulerian deformation
field.

The setup of \myreffig{fig:defoCombine}b illustrates the problem:
we start with an input shape $\imVal$ in the bottom-right quadrant,
and the two deformations (orange boxes) shown in the top row of  \myreffig{fig:defoCombine}b. 
Directions of the vectors in relevant quadrants are indicated with blue arrows.
Each deformation is assumed to move the input surface from one quadrant of the
domain to the next one for $\alpha_1=\alpha_2=1$. 
The deformation vectors in empty quadrants do not matter in this example. We can assume 
that they are relatively small, and point in a different direction, e.g., to the right.
Deformation $\velAl_1$ moves $\imVal$ left, while $\velAl_2$ moves it upwards, so
that the target configuration has $\imVal$ positioned in the top-left quadrant. 
The Eulerian representation of the deformations in this case means that the
deformation vectors moving the star to the left in $\velAl_1$ are typically not
present at the actual surface locations of the star, but in the target
quadrant.
The straightforward combination of $\velAl_1$ and $\velAl_2$ by addition yields:
\begin{eqnarray} \label{eq:alignWrong}
	\imVal(\posAl - (\velAl_1(\posAl) + \velAl_2(\posAl)) ) \ , 
\end{eqnarray}
which is clearly wrong and completely ignores $\velAl_1$ (as shown in the middle row of \myreffig{fig:defoCombine}b).

\scndRev{
Instead, it is necessary to align the deformation lookup for $\velAl_1$ with
the second deformation, as the combined deformation needs to have the right
deformation vectors in the target region where the surface should end up.
A location $\posAl$ with deformation $\velAl_1(\posAl)$ is 
not applied to $\posAl$ for the combined deformation, but it is moved by $\velAl_2$.
Instead, the deformation $\velAl_1( \posAl-\velAl_2(\posAl) )$ is the one that is applied at $\posAl$.

Thus, it is necessary to apply the deformation $\velAl_2$ to $\velAl_1$.
We do this by computing an intermediate, aligned deformation $\velAl_1'$ with:
}
\begin{eqnarray} \label{eq:alignAdvect4}
 \velAl_1'(\posAl) &=& \velAl_1(\posAl -   \velAl_2(\posAl) ) \, , 
\end{eqnarray}
which can now be combined with $\velAl_2$ as indicated in \myrefeq{eq:alignWrong} with
\begin{eqnarray} \label{eq:alignAdvect4b}
 \imDstal(\posAl)   &=& \imVal(\posAl - \alpha_1  \velAl_1'(\posAl) 
                                -  \alpha_2  \velAl_2(\posAl) ) \ .
\end{eqnarray}
This aligned combination is shown on the bottom row of \myreffig{fig:defoCombine}b. Here,
the deformation correctly combines both left and upwards motions in the top-left quadrant,
moving the input shape with a single advection step.
Note that $\alpha_2$ is only applied during the addition in \myrefeq{eq:alignAdvect4b}, and not yet in 
\myrefeq{eq:alignAdvect4}.

\revision{
This deformation alignment extends to arbitrary sequences. The process is illustrated
for three deformations below. In this case, $\imDstal$ is calculated by:
\begin{eqnarray} \label{eq:alignAdvect5}
 \imVal'(\posAl) &=& \imVal(\posAl - \alpha_1  \velAl_1(\posAl) ) \nonumber \\
 \imVal''(\posAl) &=& \imVal'(\posAl - \alpha_2  \velAl_2(\posAl) ) \nonumber \\
 \imDstal(\posAl) &=& \imVal''(\posAl - \alpha_3  \velAl_3(\posAl) )
\end{eqnarray}
which can be expressed in terms of aligned deformations as
\begin{eqnarray} \label{eq:alignAdvect6}
 \velAl_1'(\posAl) &=& \velAl_1(\posAl -   \velAl_3(\posAl)  -   \velAl_2(\posAl -   \velAl_3(\posAl) ) ) ) 
 \nonumber \\
 \velAl_2'(\posAl) &=& \velAl_2(\posAl -   \velAl_3(\posAl) ) 
 \nonumber \\
 \imDstal(\posAl)  &=& \imVal(\posAl - \alpha_1  \velAl_1'(\posAl) 
                                 - \alpha_2  \velAl_2'(\posAl)  
                                 - \alpha_3  \velAl_3(\posAl) )
\end{eqnarray}
Note that both $\velAl_1$ and $\velAl_2$ have to be aligned with $\velAl_3$ in this case.
Thus, while the last deformation of a sequence is left unmodified, previous deformations
need to be aligned by all previous aligned deformations.
This leads to our final algorithm for combining $n$ deformations: 
} %
\begin{algorithm}[h!]
 {\bf function} alignVelocity($\velV_1, ..., \velV_n$) { \\
 	$\veldst = \alpha_1 \velV_1 $\\
 	\For{i=2 \text{ to } n}{
 		\tcp{Alignment via semi-Lagrangian step}
 		$\forall \, \Vect{x}: \  \velV_{\text{tmp}}(\Vect{x}) = \veldst(\Vect{x} -   \velV_i(\Vect{x}) ) $ \\
 		$\veldst = \alpha_i \velV_i + \velV_{\text{tmp}} $\\
 	}
 	\KwRet{ $\veldst$ }
 }
\label{alg:advectAlign}
\end{algorithm}

\vspace{-0.5cm}
\scndRev{Like before, the scaling factors $\alpha_i$ are only applied when accumulating deformations 
in $\veldst$, but not when applying them to the previous deformations.}

\section{Interpolation}
\label{sec:blend}

The deformations are typically calculated for a set of inputs during a pre-processing stage.
We now explain the runtime interpolation step to generate new outputs with these deformations.
Note that we always calculate the deformations for SDF inputs, but we apply those deformations to the original data set (e.g., either a liquid 
SDF or a smoke volume) when generating an interpolated version. Thus, the inputs, denoted by $\imVbl$ in the following, are not necessarily SDFs.

\scndRev{
To perform an interpolation we require a series of input data sets $\imVbl_i$ with parameter values $\mathbf{r}_i$. %
Here, the vector $\mathbf{r}_i$ can represent arbitrary parameters of the input simulations, e.g., the xy-position of an inflow object in a plane, or its size.
The dimension of  $\mathbf{r}_i$ directly determines how many dimensions need to be interpolated to generate
an output surface.
After describing our general approach below, we will give details
of 1D and 2D versions, and then discuss a modification for liquids.
}

To interpolate a new output with chosen parameters $\tilde{\uvp}$ 
we first connect our input data points with appropriate simplices. 
Thus for a single parameter, they are connected
with line segments, and for a 2D parameter space by triangles. 
Higher dimensions correspondingly require tetrahedra or higher dimensional
simplices to discretize the volume of the parameter space.

\begin{figure}[bt] { 
	\begin{center} \includegraphics[width=1.0\linewidth]{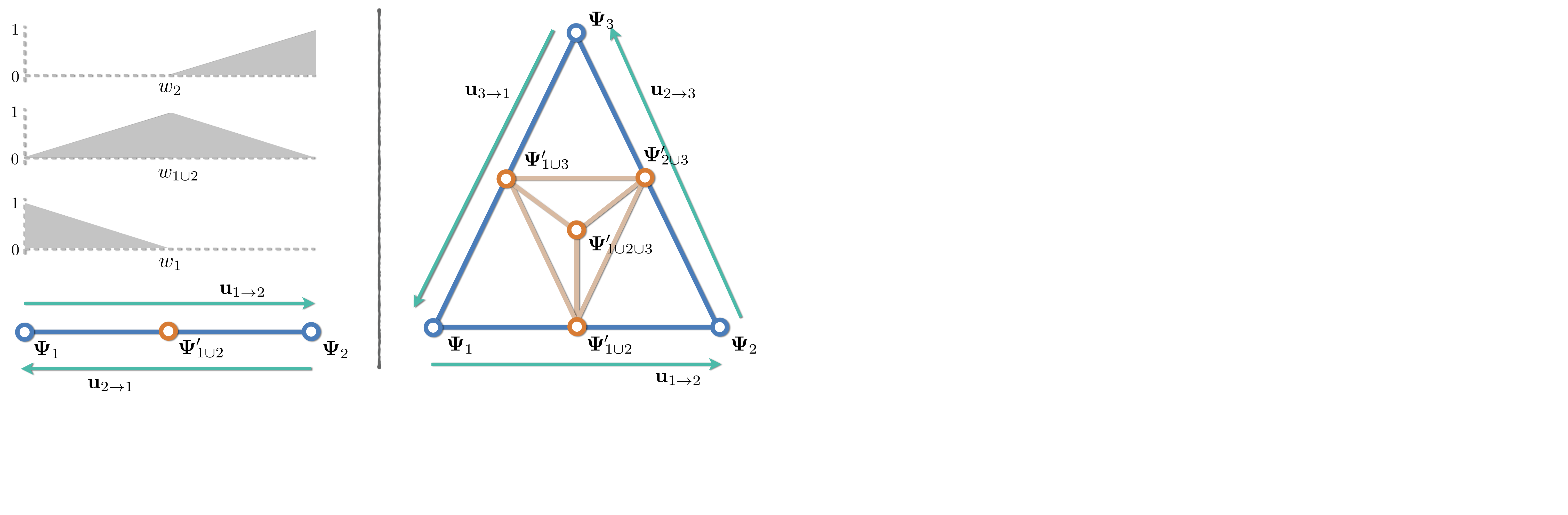}\end{center}   \vspaceImgSm
} \caption{ \label{fig:multidim} 
	Our subdivision for simplices spanning the parameter space of liquid input 
	data can be seen for a 1D (bottom-left) and 2D example (right). 
	The orange dots indicate inserted data points from SDF unions.
	The weights for each of the 
	three data points of the 1D example are shown on the left side.
} \end{figure}

We aim for an interpolation scheme that yields a smooth transition
between the inputs and that retrieves the inputs at its endpoints.
To generate an output for the parameters $\tilde{\uvp}$, we find the simplex that contains $\tilde{\uvp}$, 
and transform $\uvpt$ into barycentric coordinates. 
For this we use a regular barycentric conversion 
\vspace{-0.2cm}
\begin{eqnarray} \label{eq:blend}
\uvfunc(\uvpt, \mathbf{r}_1,...,\mathbf{r}_n) = \begin{pmatrix} \mathbf{p}' \\ 1 - p'_1 - ... - p'_{n-1} \end{pmatrix} 
\ \text{with} \ 
\\ \nonumber
\mathbf{p}' = \Big( (\mathbf{r}_1-\mathbf{r}_n) ... (\mathbf{r}_{n-1}-\mathbf{r}_n) \Big)^{-1} (\uvpt-\mathbf{r}_n)  \nonumber
\end{eqnarray} 
\vspace{-0.02cm}
that maps 
the point $\mathbf{p}$ onto the barycentric space of the $(n-1)$D simplex $S$ spanned by $\mathbf{r}_1,...,\mathbf{r}_n$.
We then deform the input data sets 
and interpolate them with the barycentric coordinates
$\uvp = \uvfunc(\uvpt, \mathbf{r}_1,...,\mathbf{r}_n) $. 
The deformations need to cover the whole simplex, but we do not require deformations for all possible
connections between the vertices of a simplex.
Instead, we order the deformations to span all sides,
and thus require $d+1$ deformations for a $d$-dimensional simplex.
This is illustrated by the green arrows in \myreffig{fig:multidim} for two examples.
Each deformation $\defo{i}{j}$ is calculated for the pair of inputs $(i,j)$ with \myrefalg{alg:mainAlg}. 

\revision{
The simplest possible case is a single parameter $x_1$ and two inputs
 $\imVbl_1,\imVbl_2$ with deformations 
$\mathbf{u}_{1 \rightarrow 2}$ and $\mathbf{u}_{2 \rightarrow 1}$. In this 
case the result $\imVbl_{o}$  is given by 
\begin{eqnarray} \label{eq:blend1d}
(x_1,x_2)^T  & =  & \uvfunc(\tilde{x}, {r}_1, {r}_2)   \nonumber  \\
\imVbl_1'  & = &  \text{advect}( \imVbl_1 , \mathbf{u}_{1 \rightarrow 2}, 1-x_1)  \nonumber \\ 
\imVbl_2'  & = & \text{advect}( \imVbl_2 , \mathbf{u}_{2 \rightarrow 1}, 1-x_2)   \nonumber \\ 
\imVbl_{o} & = & x_1~\imVbl_1' +   x_2~\imVbl_2' \ . 
\end{eqnarray}
The inputs are deformed for the desired parameter position, and the deformed
intermediate data sets $\imVbl_i'$ are then blended with the barycentric weights.
Note that the deformations to calculate $\imVbl_1'$ and $\imVbl_2'$  are applied with a factor of 
one minus the other barycentric weight in \myrefeq{eq:blend1d}. It is also worth pointing out that
it is typically not necessary to calculate the deformed inputs $\imVbl'$ for the whole 4D volume at once. 
We perform the necessary calculations only for the 3D slice of the inputs
that should be displayed next, and then scale and add these slices. This significantly reduces
memory requirements for higher dimensions.
}

This approach directly extends to 2D input parameters, by using
barycentric coordinates of triangles spanning three inputs. 
However, it is necessary here (as well as for higher dimensions), to align the
sequence of deformations. As in \myrefsec{sec:alignment} 
we have consecutive deformations that are applied to an input.
A straightforward combination would only work when both deformations contain pure translations.
For lower values, an alignment of earlier deformations using our algorithm 
from  \myrefsec{sec:alignment} is crucial to prevent undesirable motions when combining the
deformation.
Thus, for two dimensions, we
 apply deformations $\mathbf{u}_{i \rightarrow i+1}$ and $\mathbf{u}_{i+1 \rightarrow i+2}$
 to $\imVbl_i$, with weights $(1-x_{i})$ and $(1-x_{i}-x_{i+1})$ respectively,
where $\mathbf{u}_{i \rightarrow i+1}$ is aligned with \myrefeq{eq:alignAdvect4}.

\mypara{Smoke}
For smoke inputs, the deformations calculated by the optical flow step do not necessarily
conserve mass, nor do the inputs necessarily have matching total masses over time.
For the former, we calculate the total mass after applying the deformation,
and normalize it to yield the original mass of the input. This normalization
factor is calculated for each smoke deformation separately. The transition from
source to target mass is then handled by the linear interpolation. %

\revision{
\mypara{Liquids}
While this linear blend works nicely for smoke data, 
we found that the union blending technique of Stanton et al. \shortcite{stanton2014srg}
yields higher-quality results for the SDFs of liquids. 
They propose to blend via the union of both SDFs, which is especially important for 
droplets and thin structures. \scndRev{
While smaller wisps of smoke simply become transparent when they are
scaled down during interpolation, smaller features of SDFs can quickly disappear.
Blending via the union of the input SDFs preserves these small structures
much better.}

We first review the union-blending approach \cite{stanton2014srg} for two
inputs in the following, and then extend this idea to higher dimensions.
In settings where a high quality is not crucial, e.g., for real-time applications, 
uni-directional {\em nearest-neighbor} interpolation could be used. For a set of chosen
weights, the result is then calculated by deforming only the closest input, without blending other data points.
This is in line with the nearest-neighbor interpolation from Raveendran et al. \shortcite{Raveendran:2014:blendingLiquids}. 
}

\begin{figure}[t]
{ \begin{center}
	\includegraphics[width=1.00\linewidth]{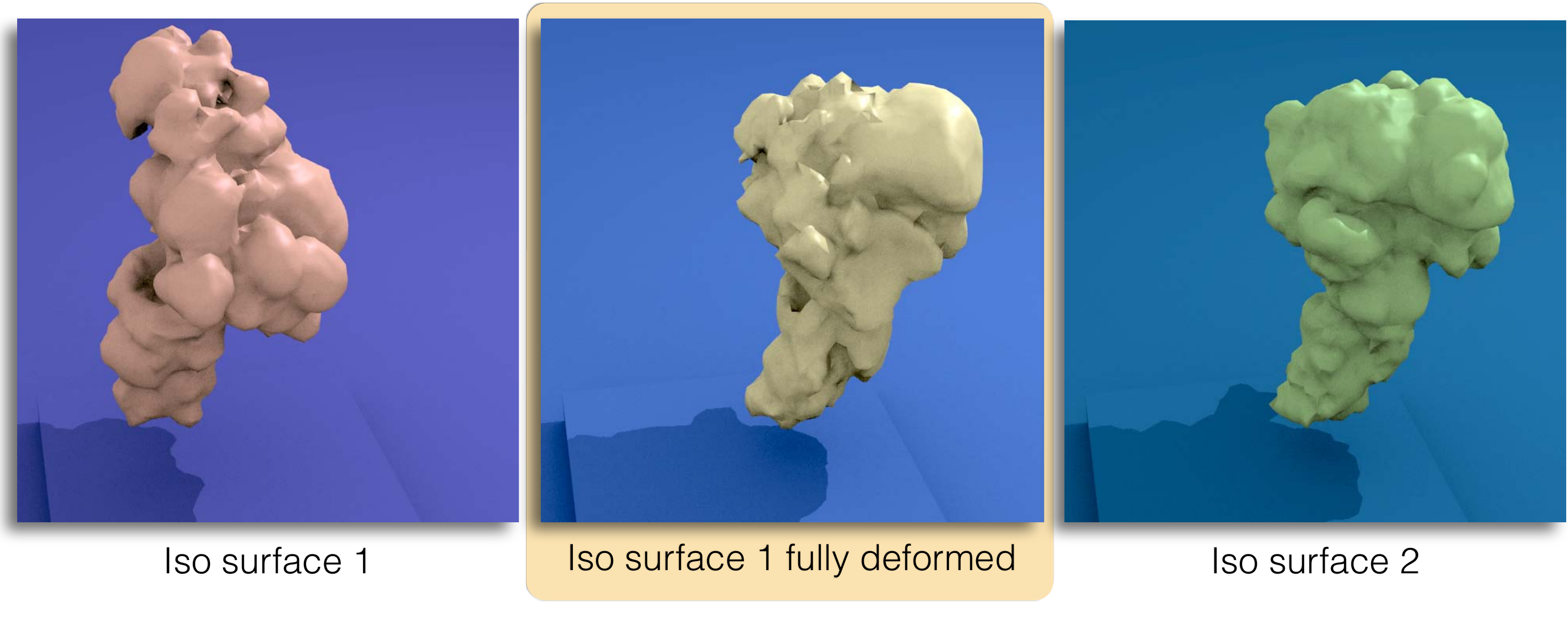}\end{center}  \vspaceImgSm
} \caption{ \label{fig:smokemesh} 
	Iso-surfaces extracted from smoke simulations. The source and target are on the left and right, respectively. In the
	middle the fully deformed source surface is shown. It closely matches the medium to large scale features
	of the target. This example uses only a single deformation, and no blending between the two inputs.
} \end{figure} 

\revision{
For all our results we use the following approach with union-blending.
For the simplest case, two inputs $\imVbl_1$ and $\imVbl_2$ from liquid simulations,
and weight $\alpha$,
this yields %
\begin{eqnarray} \label{eq:blendLiq}
\imVbl_{1 \cup 2}' & = & \text{min} (\imVbl_1' , \imVbl_2')    \nonumber   
\\   \label{eq:blendLiq_p2}
\imVbl_{o} & = &
	\omega_1 ~  \imVbl_1' + 
	\omega_{1 \cup 2} ~ \imVbl_{1 \cup 2}'  + 
	\omega_2 ~ \imVbl_2'  \ ,
\text{ with }
\\
\omega_1 &=& \text{clamp}(1-2\alpha), 
\omega_2 = \text{clamp}(2\alpha-1), 
\end{eqnarray}
where $\text{clamp}$ ensures a 0 to 1 range, and
$\omega_{1 \cup 2} = 1 - \omega_1 - \omega_2$.
The main differences to before are the different interpolation weights, and the inclusion of the temporary union SDF
$\imVbl_{1 \cup 2}'$ from the deformed inputs. This interpolation requires slightly more memory to store $\imVbl_{1 \cup 2}'$,
but in practice runs as fast as the simpler version from \myrefeq{eq:blend1d}.
} %

\revision{
We now extend our general formulation to include the union SDFs and 
propose a simple scheme to calculate the corresponding weights for higher dimensions.
For this we subdivide each initial simplex $S$ into smaller simplices $S_i$ by adding data points for the union SDFs. 
One such data point
is added at the center of each lower-dimensional simplex. Thus, for a 1D interpolation, the
union of the two inputs is added at the midpoint of the line, yielding the interpolation from \myrefeq{eq:blendLiq}.
In 2D we add three union data points along the edges of the triangle, and one union in the center. 
This subdivision scheme is illustrated in \myreffig{fig:multidim} for 1D and 2D, and easily extends to higher dimensions.
}

\newcommand{\myhalf}{\tfrac{1}{2}} 

We calculate the subdivision and the interpolation weights in terms of the barycentric coordinates of
the initial simplex.
We then check in which simplex $S_i$ the point $\uvp$ lies, and 
determine the weights for the data points involved with a suitable mapping onto the local barycentric 
coordinates of $S_i$. All other interpolation weights are set to zero.
With this scheme the weight calculation of \myrefeq{eq:blendLiq} %
can be reformulated in the following way:
\vspace{-0.1cm}
\begin{eqnarray} \label{eq:blendLiq1d}
\uvp &=& \uvfunc( \tilde{\uvp},  r_1, r_2 )
\\
\begin{pmatrix} \omega_1 \\\omega_{1 \cup 2} \end{pmatrix} &=& \uvfunc ( \uvp,  0,\myhalf) 
	\ \text{ if } \mathbf{x} \in S(0 , \myhalf) 
\nonumber \\
\begin{pmatrix} \omega_{1 \cup 2} \\\omega_{2} \end{pmatrix} 
	 &=& \uvfunc ( \uvp,  \myhalf, 1) 
	\ \text{ if } \mathbf{x} \in S( \myhalf, 1)
\nonumber
\end{eqnarray}
\vspace{-0.02cm}
The final blending of the two deformed input SDFs and their union is performed as in \myrefeq{eq:blendLiq_p2}.
\begin{figure}[bt] { 
	\begin{center} \includegraphics[width=1.00\linewidth]{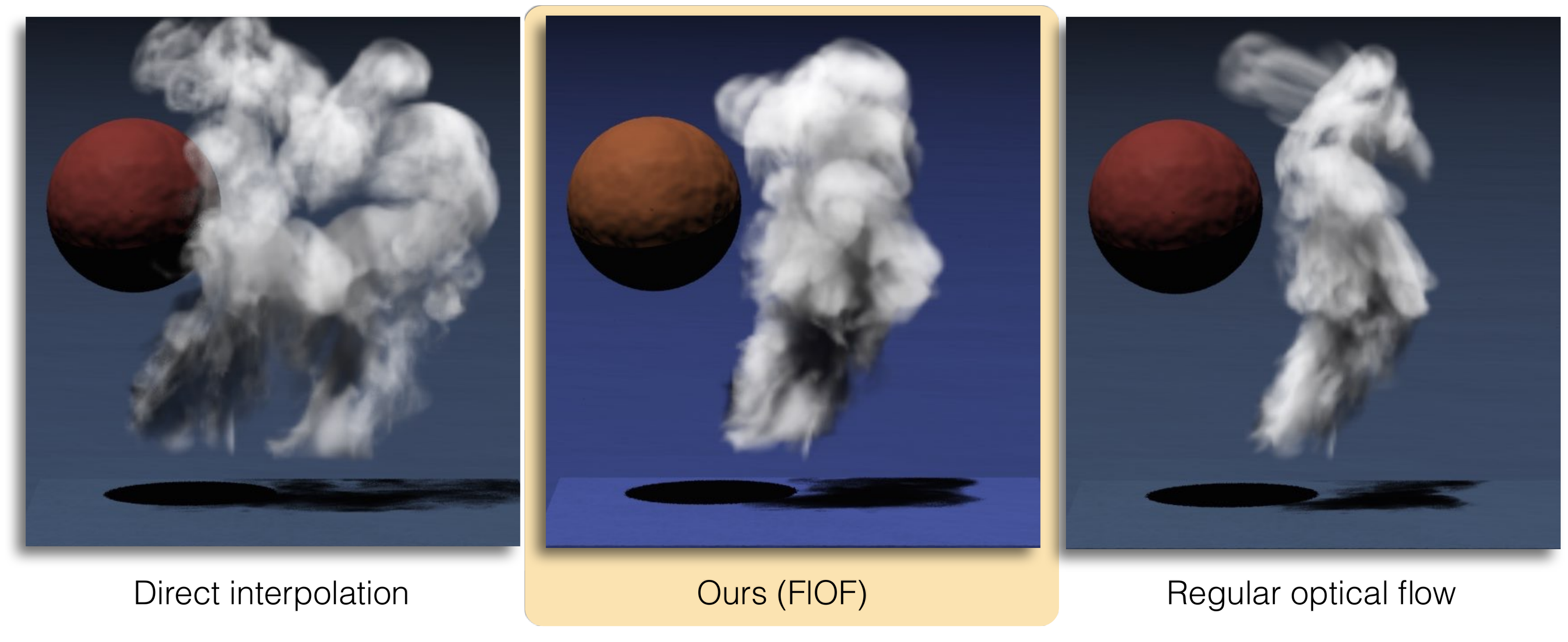}\end{center}  \vspaceImgSm
} \caption{ \label{fig:smokeCompBad} 
	From left to right: a direct blend of the two input volumes,
	a bi-directional blend with FlOF, and a bi-directional blend using a regular optical flow 
	match computed from the smoke densities.  The direct interpolation gives an
	undesirable duplication, while the regular optical flow result exhibits a clearly suboptimal 
	match. Our version in the middle leads to a very good alignment of the two inputs.
} \end{figure} 

This approach naturally extends to 2D parameter spaces.  %
An exemplary calculation of the weights for a point in the bottom-right triangle
of \myreffig{fig:multidim} (right) is 
\begin{eqnarray} \label{eq:blendLiq2dEx}
\begin{pmatrix}
	\omega_{1\cup2} \\ \omega_{2} \\  \omega_{2\cup3} 
\end{pmatrix}
 &=&  \uvfunc \Big( \uvp,  
	\begin{pmatrix}    \myhalf \\  \myhalf  \end{pmatrix} , 
	\begin{pmatrix}   0 \\  1   \end{pmatrix}  ,
	\begin{pmatrix}   0 \\  \myhalf  \end{pmatrix} 
  \Big)  \ .
\end{eqnarray}
The unions are calculated with a negligible cost from the inputs deformed by two aligned deformations.
Our approach can be used to efficiently calculate high-quality interpolations
of liquid data sets, and the barycentric calculation of subdivision and
interpolation weights simplifies calculations for higher dimensions. 

\begin{figure*}[bt] { 
	 \begin{center}
	 	\includegraphics[width=1.0\linewidth]{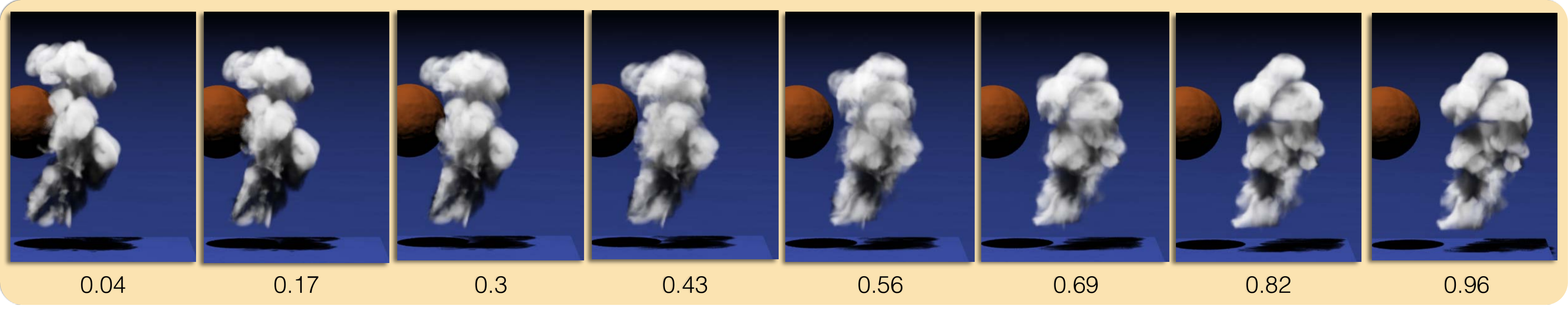}
	 \end{center}   \vspaceImgSm
} \caption{ \label{fig:smokeWltSingle} 
	The range of interpolations for the smoke setup of \myreffig{fig:smokeComp} at a fixed point in time.
	The numbers below each image indicate the interpolation weight.
} \end{figure*}

\section{Implementation}
\label{sec:impl}

Using Cartesian Eulerian grids has the advantage that 
changes of resolution are trivial, which is very useful to 
reduce the size of the optical flow solve. 
\myrefeq{eq:ofmain} is linear, and theoretically simple to solve,
but the high dimensionality of the data sets can lead to long runtimes.
In practice, we downsample the inputs and run the optical flow solve with a
resolution of around $60^4$.
To solve the linear system, we use a conjugate gradient solver, and impose $\velV=0$ at the domain boundary.
As the optical flow matrix can be ill-conditioned, we recommend using a 
diagonal preconditioner. This significantly speeds up runtimes, and performs
better than more complex preconditioners in our tests.
The deformations can then be up-sampled to generate higher
resolution outputs. 
The 5D optical flow solve itself requires about 5GB of memory with our implementation (for $60^4$).
\thirdRev{
To reduce the memory requirements when generating
the final interpolations, it is only necessary to keep a window corresponding to the maximal deformation in time in memory.
For our examples, we chose a temporal window of 20\% of the simulation length, which encompasses the
largest encountered time offsets.}
We also found that the best parameters for our algorithm are not input dependent 
and we used the set shown in \myreftab{tab:params} for all our examples.
\begin{table}[bt]
\centering
\begin{tabular}{|l|l|}
\hline Max. number of iterations & $l_{max} = k_{max} = 3$ \\
\hline Optical flow recursion threshold &  $s_{\text{max}}=10$ \\
\hline Optical flow Laplacian weight &  $\beta_{S}=10^{-3}$  \\
\hline Optical flow Tikhonov weight &  $\beta_{T}=10^{-4}$  \\
\hline SDF distance range                &   $\gamma_{\text{max}}=40$ \\
\hline Scaling of SDF inputs  & 	$\beta_{\text{image}}=-0.2/\gamma_{\text{max}}$  \\
\hline Gaussian blur kernel &  $\sigma_{\text{of}} = \sigma_{\text{proj}} = 4$  \\
\hline Projection narrow band & $\dproj=4$ \\
\hline Conjugate gradient residual threshold & $10^{-2}$ \\
\hline 
\end{tabular}
\vspace{0.2cm} \caption{List of parameters}
\label{tab:params}
\end{table}
\scndRev{%
Note that we assume a normalized cell size, thus $\Delta x = 1$, and
SDF data that is in the range $[-\gamma_{\text{max}} ~,~ \gamma_{\text{max}}]$. These 
distance values are scaled by $\beta_{\text{image}}$ for the OF solve. } 

We evaluated a variety of options for each of the algorithmic components of the
optical flow. E.g., different methods to interpolate, blur and advect the surfaces are 
imaginable. %
We were surprised to find that some of the recommended best
practices for optical flow \cite{ofsecrets2014} did not lead to better deformations.
For example, median filtering and cubic interpolation did not reduce the error.
Especially the latter led to increased error measurements for some tests.
Likewise, sub-stepping the advection for different CFL conditions, or using
higher order advection schemes \cite{Selle:2008:USM} did not pay off.
Our intuition here is that the synthetic data of the SDFs is noise-free and
smooth. The alternative components mentioned above typically introduced
high-frequency details that in turn required more smoothing in later stages.
As a result, we use simple first-order semi-Lagrangian advection, linear
interpolation, and a regular Gaussian blur.

The optical flow algorithm by default puts more emphasis on larger regions of the
surface (the energy is minimized in a least squares fashion, equally weighting every point in space). 
To make sure the first frame of the simulation is registered
properly, we repeat it in time (5 time steps for our examples).
Furthermore, we leave an empty region of ten percent at all sides of the input, to give the optical
flow the chance to freely push surfaces near the sides in- and outward.

We also noticed that the extracted time slices for liquid SDFs can exhibit slight flickering.
To alleviate this, we apply a temporal filter, for which we found the union of two adjacent extracted time slices
to be the most efficient choice. \scndRev{ Alternatively, the use of higher-oder interpolations
in time would be possible here.}

\section{Results}
\label{sec:results}

In the following we demonstrate our matching approach with several
example inputs from smoke and liquid 
simulations.
\scndRev{All timings were measured using an Intel Core i7 with 4 GHz,
and were generated with {\em mantaflow} \shortcite{mantaflow}.}

\mypara{Smoke}
As a first test case we consider two buoyant smoke clouds. The clouds have
different initial positions, and only one of them directly interacts with a
spherical obstacle in the scene. This leads to distinctly different shapes and
motions of the two clouds. A successful $0.5$ interpolation of these clouds
with our method is shown in the middle of \myreffig{fig:smokeComp}.
Generating the smoke volumes took only $0.56s$ per frame on average for this
example. This extraction timing (as well as the following)
includes all calculations necessary to extract the final volume to
be displayed, excluding disk I/O.

\fourthRev{We have used an iso-level of $10\%$ of the maximal density to define the surface, and this
value was used for all other smoke inputs below.}
The iso-surfaces extracted in this way do not conserve volume, and can change
significantly from one frame to the next (\myreffig{fig:smokemesh}, left and right). Despite these difficulties our
algorithm robustly recovers a match between the inputs, and the final
deformation (\myreffig{fig:smokemesh}, middle) matches the target shape very
well.
\begin{figure*}[t] { 
	 \begin{center}
	 	\includegraphics[height=104pt]{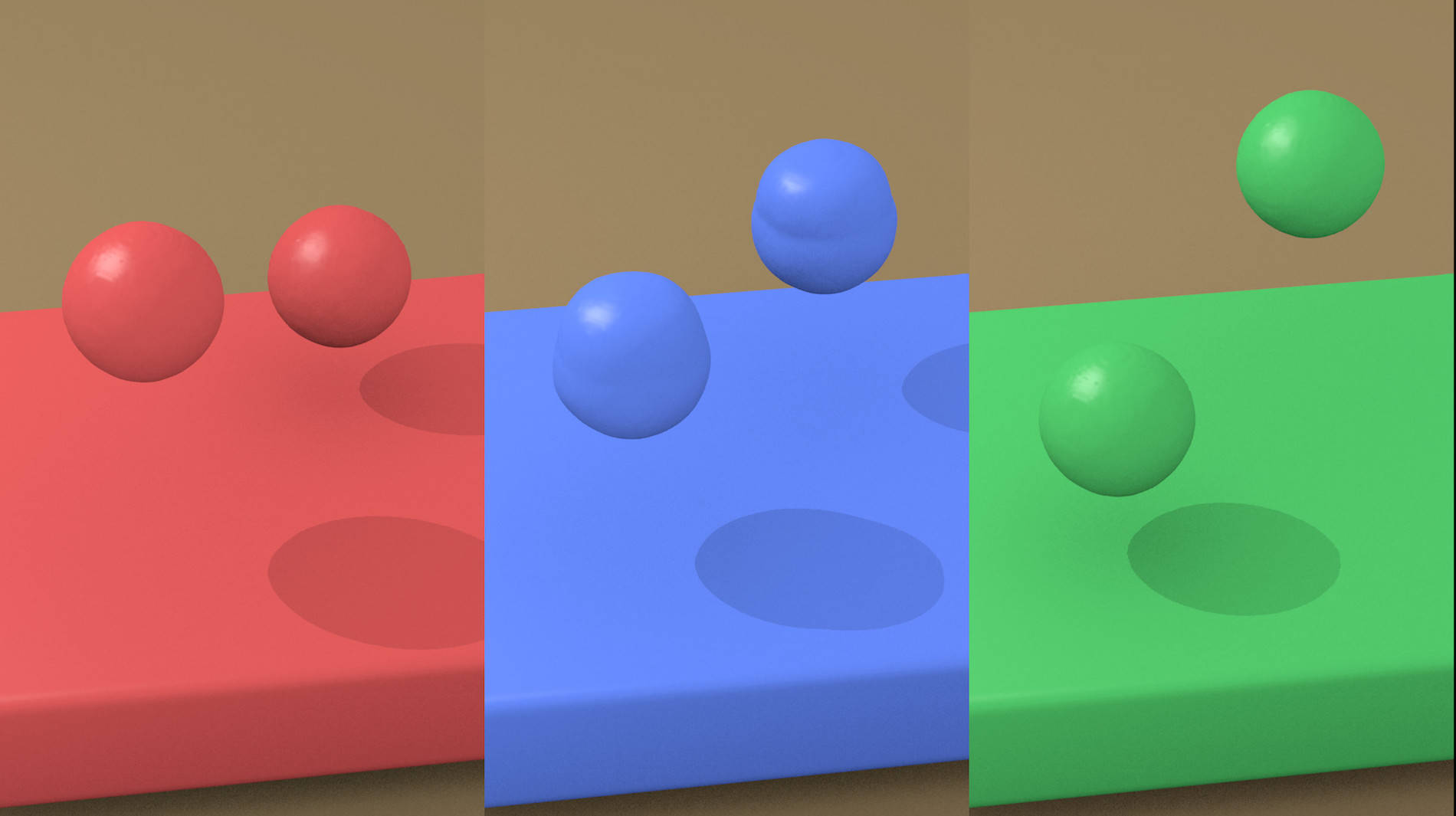}
	 	\includegraphics[height=104pt]{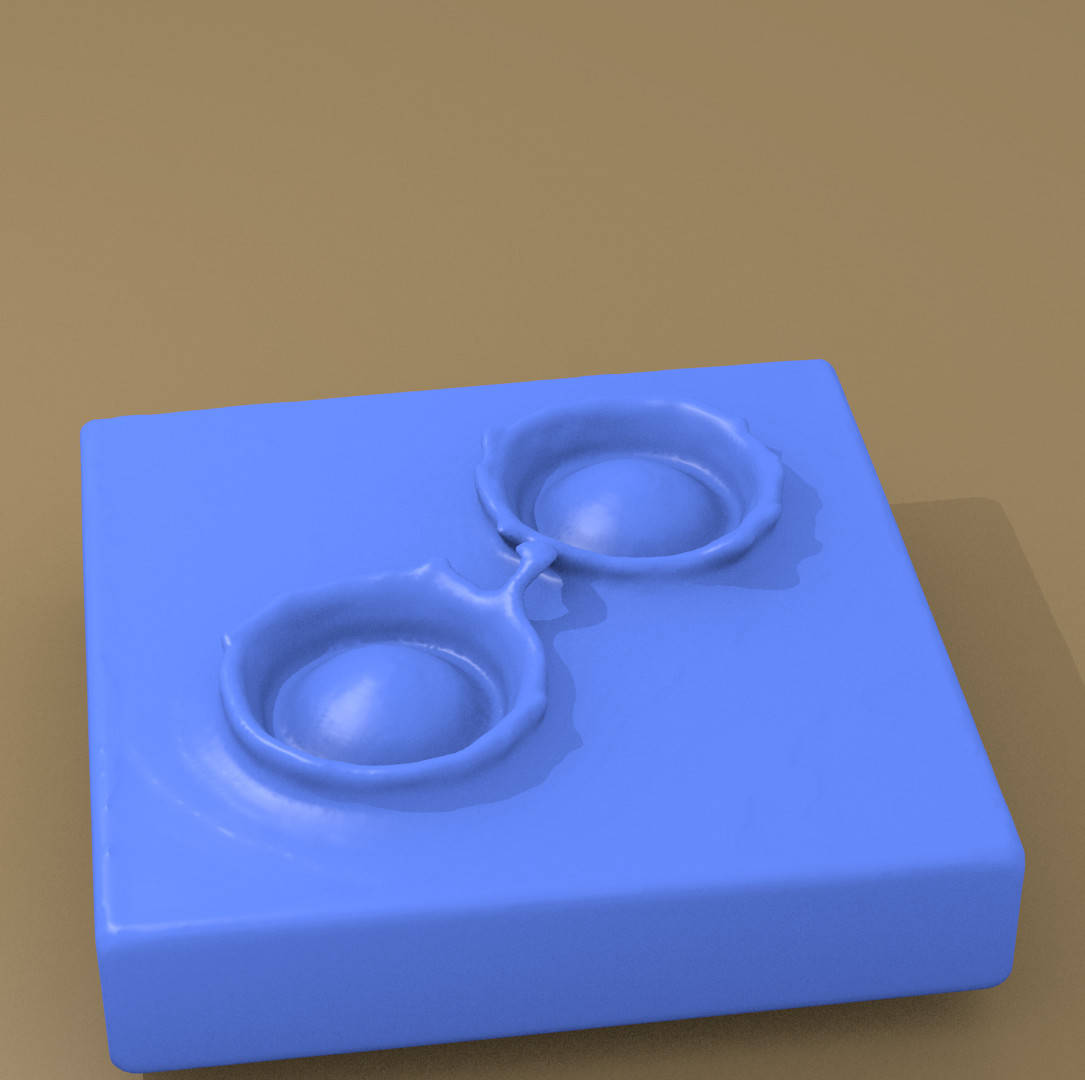}
	 	\includegraphics[height=104pt]{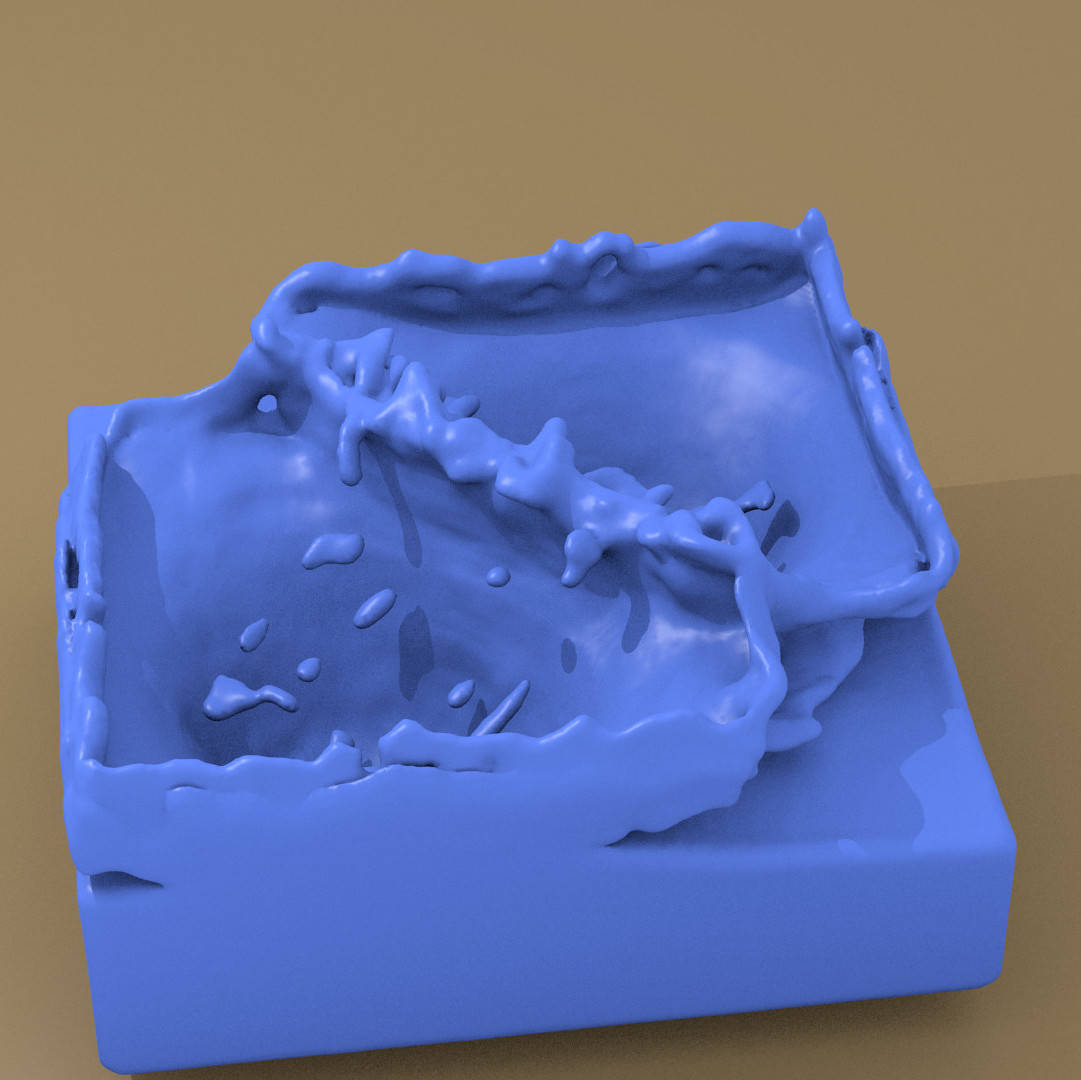}
	 	\includegraphics[height=104pt]{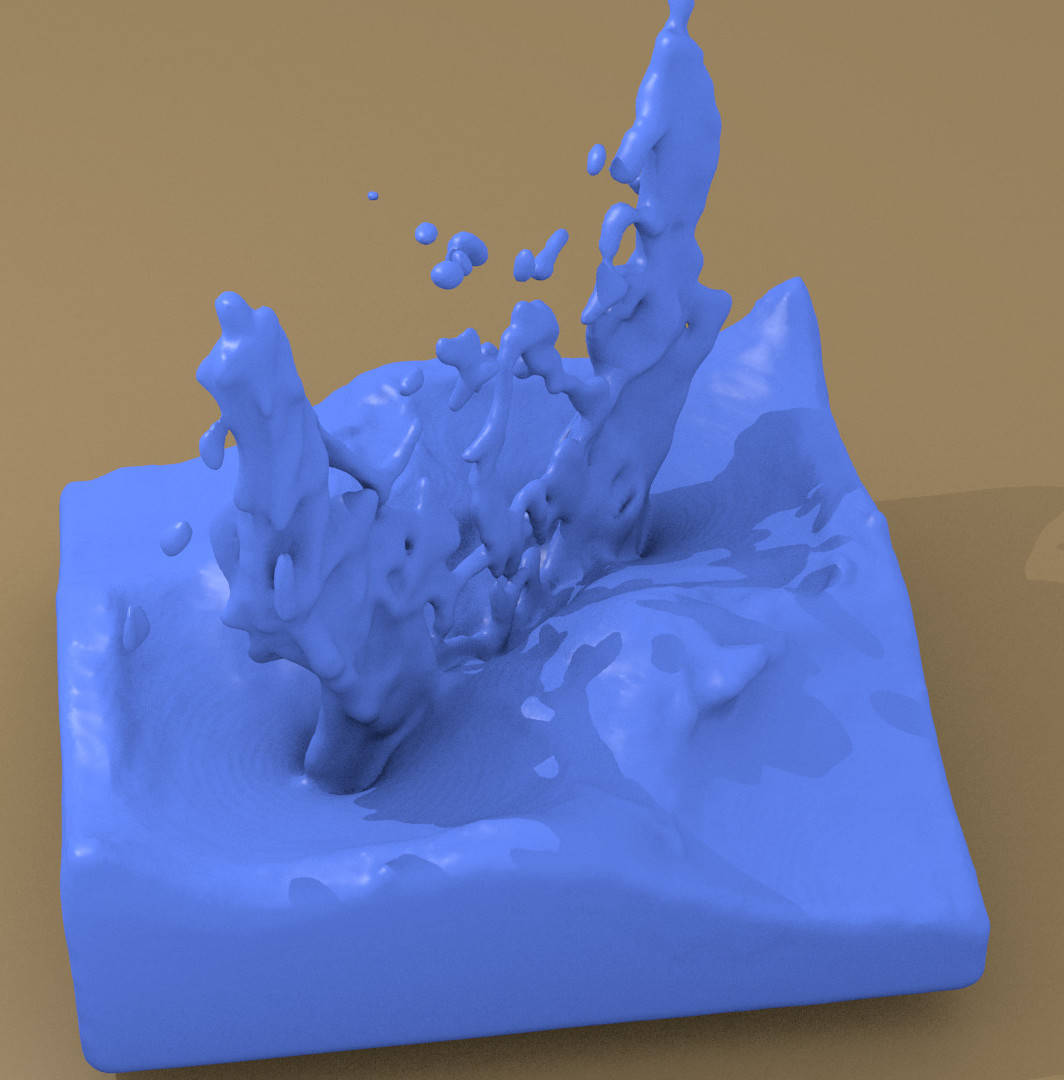}
	 \end{center}   \vspaceImgSm
} \caption{ \label{fig:kdrop} 
	An interpolation of two liquid simulations. The inputs (in red and green on the left), have different initial drop positions. Our
	$0.5$ interpolation (in blue) aligns them in space and time so that the impact of the drops coincides. The three images on 
	the right show later frames of our interpolated version.
} \end{figure*} 
\begin{table}[bt]
\centering
\begin{tabular}{|c|c|c|c|c|c|}
\hline Fig.  & \multicolumn{2}{c|}{Simulation} &  \multicolumn{3}{|c|}{Deformation}  \\
\hline       &  Res. & Time &  Res. & Mem. & Time \\
\hline \ref{fig:smokeComp}    &  $200^2 \mycdt 300 \mycdt 150  $  &  24m 2s      &   $50^2 \cdot 75^2 $  & 4.96  & 6m 16s              \\
\hline \ref{fig:smokeCompBad} &       ''                          &      ''     &    $50^2 \cdot 75^2 $ & 2.80 & 1m 27s$^{(+)}$      \\
\hline \ref{fig:kdrop}        &  $200^3  \cdot 450         $      &  30m 21s    &   $54^3 \cdot 81 $    & 4.56 & 4m 24s             \\
\hline \ref{fig:streamComp}   &  $240^3  \cdot 360          $     &  25m 12s    &   $54^3 \cdot 81 $    & 5.35  & 4m 53s             \\
\hline \ref{fig:liqsmComp}    &  $300^3  \cdot 450          $     &   34m 16s    &   $50^3 \cdot 75 $   & 2.81 & 3m 07s             \\
\hline
\end{tabular}
\vspace{0.2cm} \caption{Timings for all example scenes} \label{tab:timings}
\end{table}
The whole range of the complex interpolations calculated by our method is shown
in \myreffig{fig:smokeWltSingle} for a static frame. Resolutions and timings for this
and the following simulations can be found in \myreftab{tab:timings}.
\scndRev{
Here, memory requirements are given in GB, and 
Timings %
are averaged values across a full simulation or FlOF solve.}

\myreffig{fig:smokeCompBad} shows a comparison of our method with other approaches
using the same smoke inputs. 
On the left is a direct blend of two input volumes, which yields a clear duplication
of the smoke volumes. On the right, we applied a state-of-the-art hierarchical 
optical flow \cite{meinhardt2013horn} to the two smoke volumes. 
The resulting deformation pushes the volumes in the right
direction, but gives very undesirable stretching in space and time. 
\scndRev{
For this optical flow solve, the CG residual 
threshold in \myreftab{tab:timings} was lowered by 50\% to ensure the match is as good as possible,
and it uses identical input data as \myreffig{fig:smokeComp}.}
Our result, shown in the middle of \myreffig{fig:smokeCompBad}, successfully aligns the two clouds, giving a single
rising column of smoke with an intermediate shape.
Smaller wisps of smoke around the main cloud are moving along at no extra cost.

\mypara{Liquid}
\myreffig{fig:kdrop} demonstrates the FlOF algorithm and our interpolation scheme
for a liquid example. The two inputs are simulations with the initial position of the drops
interchanged. In the inputs the drops impact at different times, and the resulting splashes are asymmetric. Our 
deformation correctly retrieves a simultaneous impact for an interpolation with weight $0.5$.
A similar setup was simulated by Raveendran et al. \shortcite{Raveendran:2014:blendingLiquids} using a
significantly less detailed simulation that did not contain any drops or thin sheets. The accompanying
video shows a direct comparison to illustrate the difference in surface complexity between this
version and ours. 
In contrast to the explicit detection and removal of small features proposed for the ICP-based matching \cite{Raveendran:2014:blendingLiquids}, 
our Eulerian SDFs contain averaged quantities in a broad band around the surface
that we can match robustly.
\thirdRev{
Additionally, our volumetric deformation fields can be applied to small-scale features near the surface
without additional work. For a mesh-based approach, an additional extrapolation step would be required to extend
the deformations into the volume. However, the ICP-based approach recovers the pure translational initial
configuration of the two drops with higher accuracy.}

For this example the calculation of the $240^3$ outputs with union blending (\myrefsec{sec:blend}) took 
only $0.64s$ per frame on average. %
\thirdRev{
The largest deformations for this example have a magnitude of more than $150 \Delta x$ in 4D space.
This illustrates the large distances captured by our deformations.}

\begin{figure*}[bt] {  \vspaceImgSm
	\begin{center}  \includegraphics[width=0.95\linewidth]{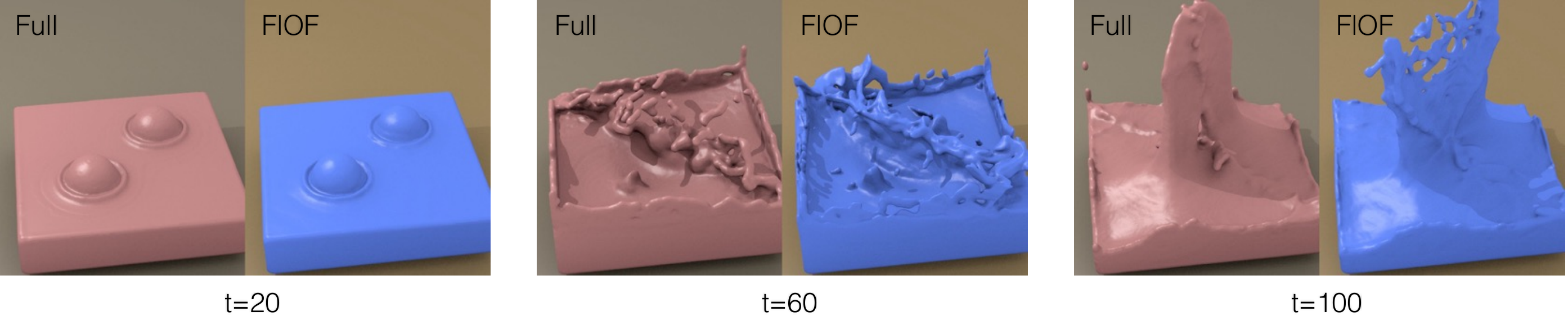}
	\end{center} \vspaceImgSm  \vspace{-0.21cm}
} \caption{ \label{fig:kdropGt} 
	\fourthRev{Several frames comparing our interpolated result at $0.5$ with a full simulation with a corresponding initial setup (i.e. drops
	at same height). Each image shows the full simulation on the left, ours on the right. While small scale structures deviate, our interpolation recovers the large-scale structures.}
} \end{figure*}

\thirdRev{For this liquid setup and the smoke setup of \myreffig{fig:smokeComp},
comparisons between our interpolated result and a new simulation at the intermediate position
can be found in the supplemental video (several frames for the two-drop liquid test case are shown in \myreffig{fig:kdropGt}). 
As our method generates the output based on simulations
with a different parametrization, it does not yield small-scale details that are identical to those of a full simulation, 
but it faithfully captures the behavior of the larger scales.}

\mypara{2D Parameters}
\revision{
To demonstrate interpolations within a 2D parameter space we have simulated
the liquid setup shown in \myreffig{fig:streamComp}.
A liquid inflow on the left side is positioned at various depths and heights, some of
which cause the liquid to hit the obstacle on the right wall, while other inflow positions partially hit or completely miss it.
This leads to strongly varying splashes and waves in the inputs, the full sequences of which
can be found in the supplemental materials. We use 7 different input surfaces, yielding
interpolations with 6 different triangular simplices covering a large space of fluid behavior.
We calculated 12 deformations to cover this parameter space.
As our algorithm works without requiring any user input, all deformations were generated automatically with the same set of parameters.
Example outputs using a variety of deformation combinations
can be found in \myreffig{fig:teaser} and \myreffig{fig:streamComp}.
}

For this example, the obstacle on the right wall leads to an increased difficulty for our algorithm.
We do not use this prior knowledge about the scene geometry for matching, and thus it is not guaranteed
that the deformations do not push parts of the surface into the obstacle. 
Our results indicate what can be achieved without incorporating this information into the solve
for six different simplices spanned by two deformations each.
The large number of small-scale drops and splashes around the obstacle that are successfully matched
and deformed highlight the complexity of the inputs our algorithm can deal with. \scndRev{However,
in several instances flickering artifacts and surface break-up is noticeable. The following section
will outline in more detail which parts of our algorithm are causing these.}

The interpolations in a 2D parameter space led to surface extraction times of $0.87s$ per frame on average.
As the operations involved purely consist of simple operations on regular grids, there is significant potential
for optimization with parallel processing, e.g., by using GPUs (which were not used in our implementation).

\mypara{Smoke and Liquid}
Finally, the example of \myreffig{fig:liqsmComp} 
matches an input of a falling drop of liquid with a blob of heavy smoke.
The difference of the two phenomena demonstrates how
our method can cope with challenging inputs: the drops have very different falling speeds,
and while the liquid leads to splashes and waves, the smoke buoyancy gives rise to
many complex swirling surfaces. 
These buoyant swirls translated into a complex iso-surface, that was successfully 
matched to the liquid one by our algorithm.
Once the match is calculated (treating both inputs as density data), we can easily interpolate any
in between behavior. 
This controllable transition of behavior is clearly beyond the scope of a
regular fluids solver. 
Our interpolation also conveniently blends between the differing viscosities
of the FLIP and Eulerian-only simulations, and illustrates that our method
is agnostic to the type of input: arbitrary parameters could be varied,
from scene geometry to physical parameters. 

\vspace{0.5cm}
\section{Applicability and Limitations}
\label{sec:limit}

Our approach clearly does not work for all inputs: if the two surfaces share no similarities,
we can compute a deformation, but 
the resulting interpolations will yield unexpected or undesirable
results. 
As it is difficult to explicitly specify a region
of applicability, we present a series of 
tests below to illustrate in which cases our method
yields the expected results, and where it is likely to fail. 
For all of the failure cases below a straightforward fix is to
insert additional data points, but we will discuss alternative directions
for future extensions where appropriate. %
We distinguish limitations of computing deformations with our FlOF algorithm,
and limitations arising from the subsequent interpolation.
Full animations of inputs and outputs for all cases below can be found in the supplemental video. 
Unless specified otherwise, the tests below use data-sets with a resolution of $64^4$ and deformations
of $50^4$.

\mypara{Deformation}
Below, we will show the results of applying a single deformation to sequences
which are as simple as possible,
such that the quality of the resulting deformation can be evaluated visually.

\begin{figure}[bt] { 
	 \begin{center} 
	 	 \includegraphics[width=1.0\linewidth]{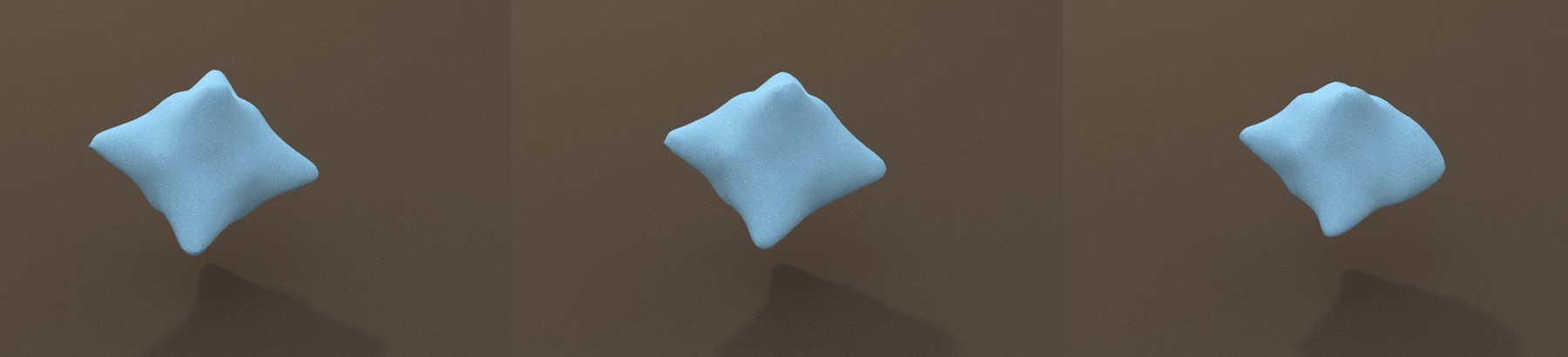}
	 \end{center} \vspaceImgSm
} \caption{ \label{fig:failShape} 
	\scndRev{
	Matching a single star-shaped surface with increasing distances. From left to 
	right: a distance of 10, 20, and 30 cells, respectively (with an interpolation
	weight of $0.4$). For the largest distance 
	the tips of the star are not fully recovered anymore. }
} \end{figure}

\begin{figure*}[bt]
{ \begin{center}   \vspaceImgSm
	\includegraphics[width=0.96\linewidth]{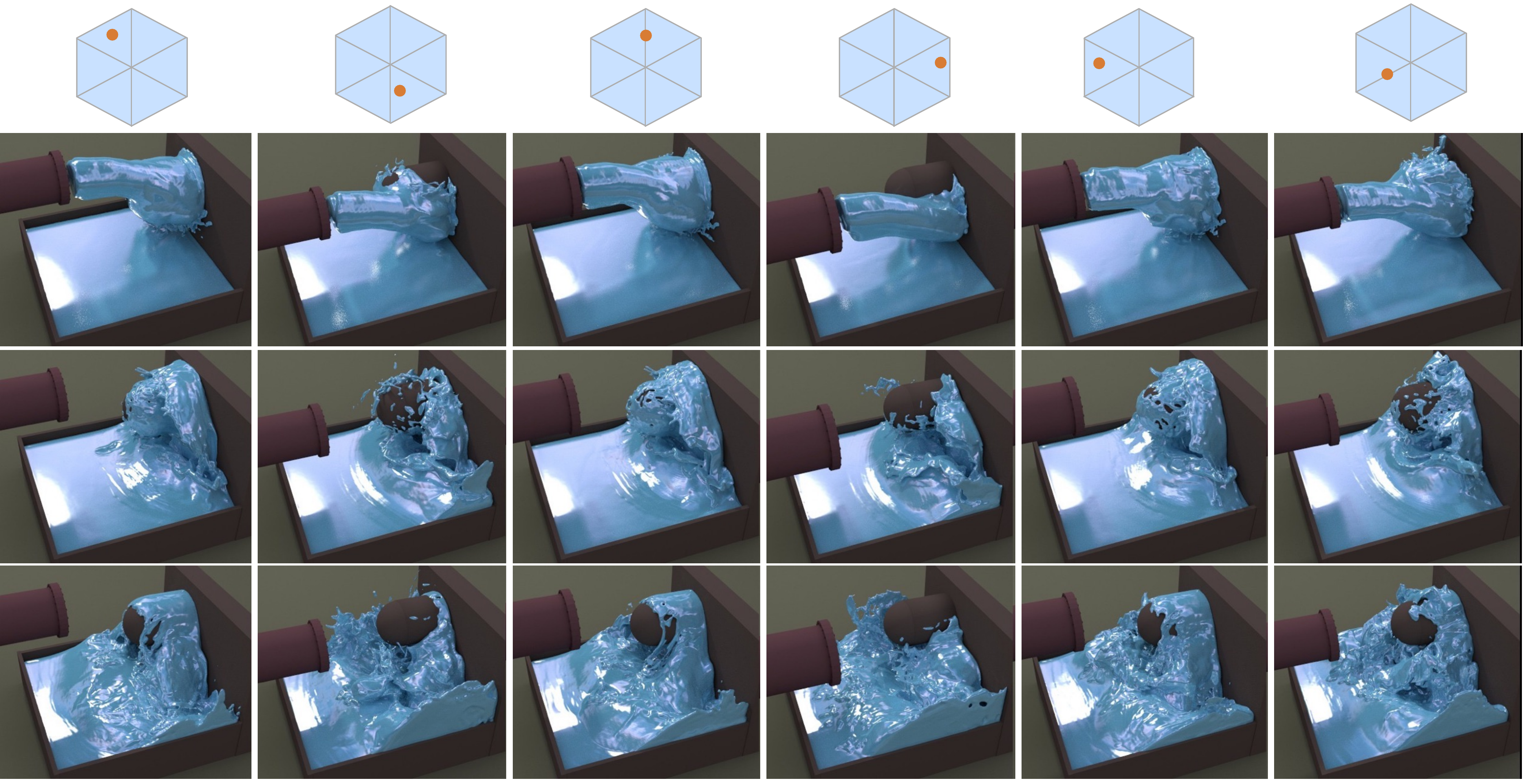}\end{center}  \vspaceImgSm
} \caption{ \label{fig:streamComp} 
	Six examples of interpolations at different points of a 2D parameter space  (shown at the top) for a liquid setup.
	The outputs exhibit strongly differing behavior, with distinct splashes and wave motions depending on where 
	in the scene the inflow stream hits the obstacle and pool.
} \end{figure*}

\thirdRev{
The matching with our algorithm is inherently based on closest distances in
four-dimensional Euclidean space. This can lead to unexpected
correspondences in certain situations.
E.g., %
for two spherical shapes that are more than their radius apart, a closest distance
will create a correspondence between two opposing sides of the spheres. 
For such a case, we have to rely on the
smoothness constraint across multiple scales in the optical flow solve to match
the correct sides of each sphere. As a consequence, our method does not always recover
an ideal rigid translation for large distances of the inputs.
This effect is noticeable for the spherical drop of \myreffig{fig:kdrop}, and the
cylindrical stream of \myreffig{fig:streamComp}. %
Additionally, if the inputs share little or no similarities, the deformations will try 
to match pieces in proximity, but not necessarily the whole target shape.  
Note that the time step of the input data also influences the relative scaling
of spatial and temporal distances.

We illustrate the behavior of our algorithm in \myreffig{fig:failShape}, 
where we match two identically moving star shapes
with increasing spatial distances.}
As the distance grows to more than ca. $25\%$ of the domain size, 
the deformation has trouble resolving the full shape of the target.
As can be seen in the supplemental video, the final shape is matched quite well,
but the non-even distribution of surface points (i.e., a non-rigid deformation of the input)
leads to a loss of features for interpolation weights less than one.

\begin{figure}[bt] { 
	 \begin{center} 
	 	 \includegraphics[width=1.0\linewidth]{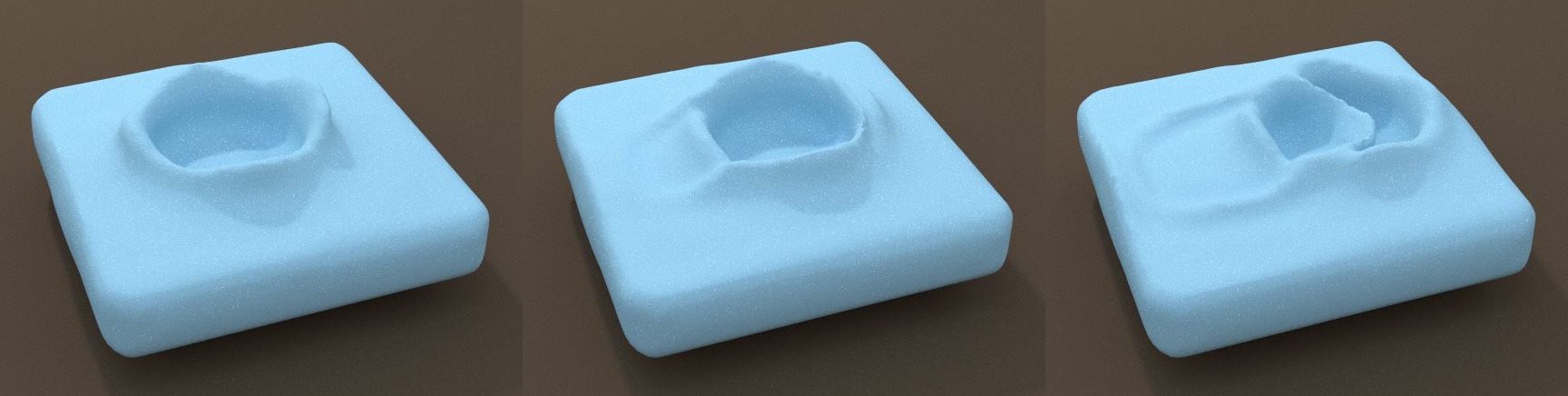}
	 \end{center} \vspaceImgSm
} \caption{ \label{fig:failBoxes} 
	\scndRev{
	The resulting surfaces for deformations computed with two falling drops with increasing distances (20, 30, and 40
	cells from left to right).  For a distance of 40 cells (right), the source splash is
	still visible after applying the deformation. }
} \end{figure} 

Moving to slightly more complicated inputs, the surfaces shown
 in \myreffig{fig:failBoxes} exhibit topology changes when a drop
hits a basin of liquid. With growing distances between the inputs,
the deformation starts to have difficulties
resolving the ambiguous surfaces around the time of impact.
For both cases above it could help to
automatically detect, and match feature points (such as a topology change).
These matches could then be enforced as constraints in the FlOF solve to 
better match important features of the inputs.

\begin{figure}[bt] { 
	 \begin{center}
	 	 \includegraphics[width=1.0\linewidth]{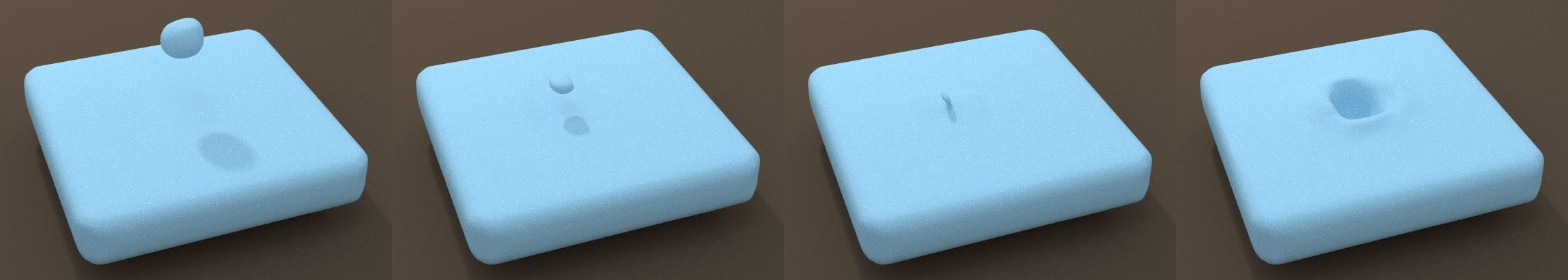}
	 \end{center} \vspaceImgSm
} \caption{ \label{fig:failTimestep} 
	\scndRev{
	A deformation of a small, fast moving drop over time. \thirdRev{While the main motion
	is recovered, the surface can start to break up during the impact for the fast moving data sets.} }
} \end{figure}

Overall, our method is quite robust w.r.t. matching small components, and fast
moving objects, as long as they are represented at the resolution
of the FlOF solve. This is illustrated with the setup of \myreffig{fig:failTimestep}, 
where the size of a falling drop is continually
decreased while the time-step of the simulations increases.
For each different set of parameters we match 
two 4D surfaces where the drops have a fixed spatial distance of 20 cells. %
The translational component is recovered even 
for small and quickly moving drops, the most difficult part being the topology change.
We have used sizes of 8, 5, and 3 cells, and time steps of 2.0, 2.75, and 3.25, respectively (for
reference, the drop example of \myreffig{fig:failBoxes} uses a size of 10 and a time step of 0.85).
The version with a time step of 2.75 is shown in \myreffig{fig:failTimestep}.

\begin{figure}[bt] { 
	 \begin{center}
	 	\includegraphics[width=1.0\linewidth]{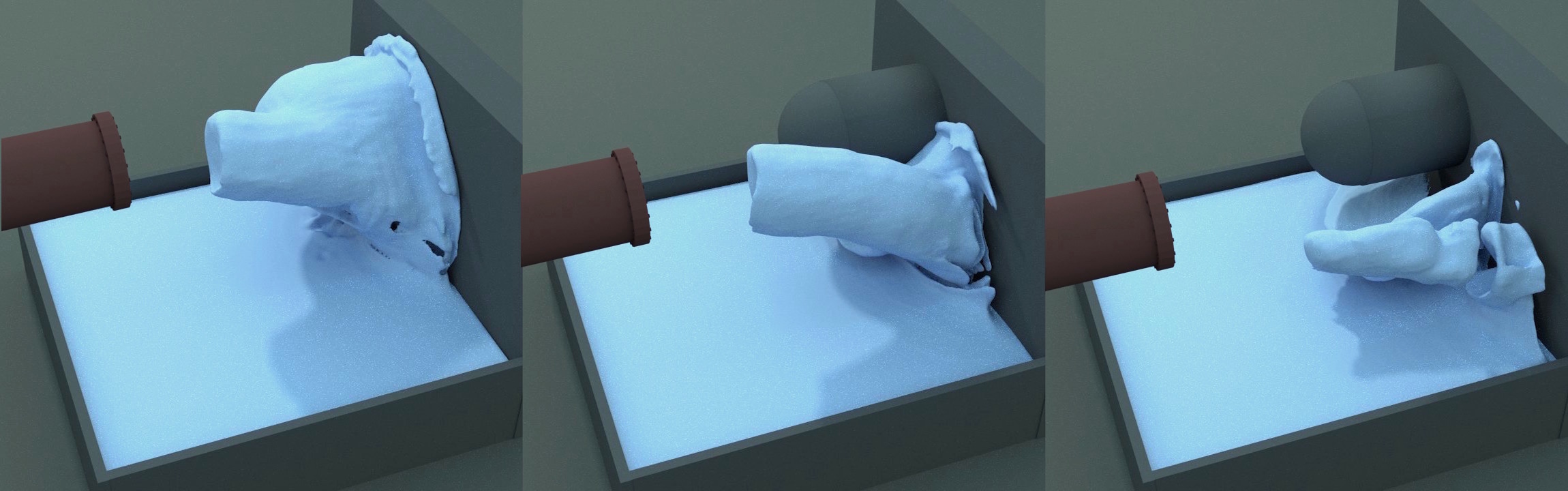}
	 \end{center} \vspaceImgSm
} \caption{ \label{fig:failure} 
	\scndRev{
	A more complex failure case: we match a
	simulation with an inflow in the back %
	with one that is moved further
	and further to the front. The images above show stills at the same time for an interpolation weight of 1.
	The target simulations had inflow shifts of $18,32$, and $46$ percent of the domain size (from left to right). }
} \end{figure}

A more complex example of a failure case for our matching algorithm is shown in
\myreffig{fig:failure}. This setup is based on the liquid stream from
\myreffig{fig:streamComp} (parameters can be found in \myreftab{tab:timings}). 
\thirdRev{A large distance between source and target shape results
in a deformation that does not fully retrieve the target shape when it is applied.
This is particularly visible for the cylindrical stream of the inflow, in the right
image of \myreffig{fig:failure}: the cylinder is only partially present in the deformed output.

As a rule of thumb, distances larger than ca. $25\%$ of the domain size can
lead to sub-optimal deformations with our algorithm. 
This estimate is partially a result of our implementation, which coarsens the optical
flow solve until a minimum resolution is reached ($s_{\text{max}}$ in \myrefalg{alg:mainAlg}).
With a different coarsening strategy, e.g., enlarging the domain along with the coarsening, potentially larger distances could be successfully matched.
A second aspect to take into account are relative distances in the data sets.
As optical flow matches surfaces in a nearest-neighbor fashion, a match across a large
distance is only made if no suitable surface in closer proximity is found. For the data sets
shown, we found that the 25\% rule prevented most ambiguities for the large-scale features of the surface.}

\thirdRev{
Another difficulty in \myreffig{fig:failure} is the presence of the obstacle on the right, which 
leads to substantially different 
liquid motions and surface shapes for the splash against the wall. For the data sets
with large distances this leads to flickering and dissolving surfaces.
While using higher resolutions for the output reduces the chance of surface flickering,
a better way to prevent these artifacts would be to introduce
a third data-point in between the two existing ones, reducing the relative distances
of the inputs. 
This would help to establish the desired correspondences based on closest distances,
and could potentially be done automatically
based the final error measurement after the FlOF solve.}

\mypara{Interpolation}
Our interpolation step combines two or more deformations to produce an output.
While the previous paragraph illustrated the behavior of a single
deformation, we now show interpolations using two deformations
and two inputs sequences (with per frame resolutions of $192^3$).

\begin{figure}[bt] { 
	 \begin{center}
	 	 \includegraphics[width=1.0\linewidth]{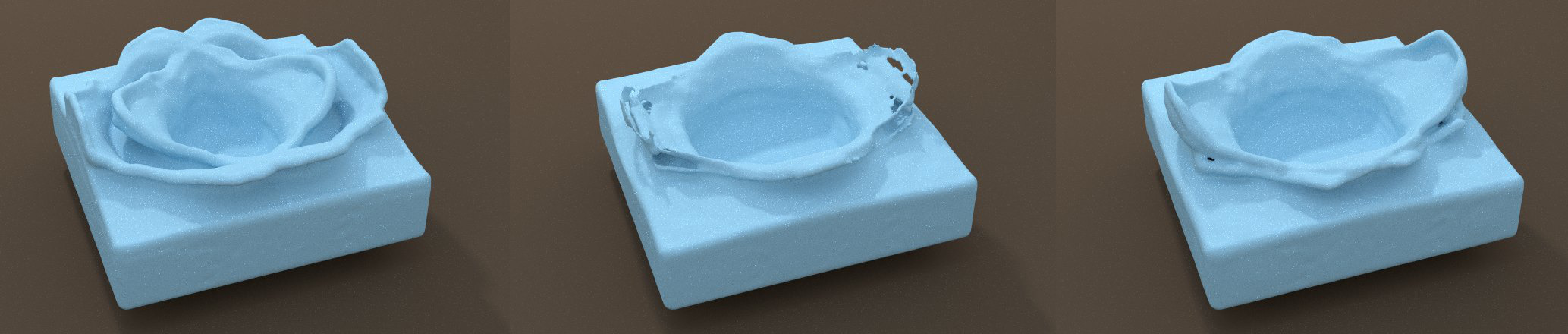}
	 \end{center} \vspaceImgSm
} \caption{ \label{fig:failIpol} 
	\scndRev{ Three different versions of an interpolation of two falling drops for weight 0.5. From left to right:
	a union of both inputs without deformation, a linear interpolation with deformation, and ours. }
} \end{figure}

\myreffig{fig:failIpol} illustrates the advantages and disadvantages of the union blending from
\myrefsec{sec:blend}. On the left a simple union of the undeformed inputs 
can be seen for reference. In the middle, a deformation is applied with strength $0.5$ to both inputs, and the deformed
surfaces are averaged to produce an interpolation at the midpoint. While the splashes align, the thin sheets
break up, leading to flickering effects during animations. The version on the right uses the same deformation, but
interpolates both surfaces with the union blending approach. This significantly reduces the 
likelihood of thin structures breaking up. 
However, %
it can lead to duplicate surfaces, as can be seen, e.g., on lower left side of the thin sheet of 
\myreffig{fig:failIpol} (right).
For smoke simulations, such misalignments can lead to 
ghosting artifacts at the sides of clouds. %

\begin{figure}[bt] { 
	 \begin{center}
	 	 \includegraphics[width=1.0\linewidth]{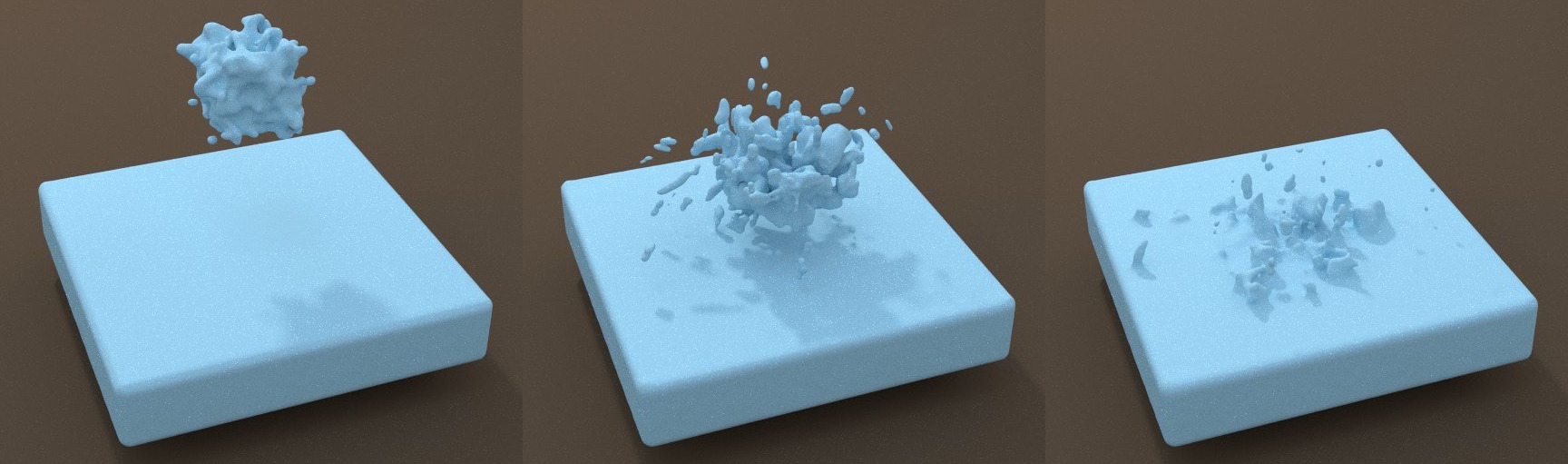}
	 \end{center} \vspaceImgSm
} \caption{ \label{fig:failSparse} 
	\scndRev{
	Several frames over time 
	of an especially difficult matching problem: a collection of randomly moving 
	small droplets. }
} \end{figure}

A particularly tough case for matching and interpolation is shown in \myreffig{fig:failSparse}:
here two sets of randomly moving drops are registered with each other (with a spatial distance
of 20 cells). 
In contrast to \myreffig{fig:failTimestep} the individual drops are not well
represented anymore for the optical flow solve.
The accompanying animation shows how an increasingly strong motion
leads to difficulties resolving the different features of the two inputs, and results in
flickering motions of the droplets.
While our algorithm still recovers the overall translation, the independently moving drops
cannot be resolved.
A similar effect is noticeable in some of the interpolations of
\myreffig{fig:streamComp}.
These artifacts could be reduced by increasing the resolution of the FlOF solve,
or by switching to a particle representation for the drops.
Additionally,
our algorithm could be used in conjunction with one or more post-processing
steps, e.g., to generate spray particles and other secondary effects based on the
interpolated surfaces.

Note that combining deformations is easier in a
Lagrangian setting, e.g. for meshes. However, we believe that the gains in
robustness and surface quality outweigh the additional computational cost for
aligning Eulerian deformations.

\section{Discussion} 
\label{sec:future}

Our approach relies on a reasonable iso-surface thresholding when working with a smoke simulation. 
However, we had no problem selecting a suitable threshold for our tests. 
For scenarios with lots of uniform densities it would also be possible to
invest more computational work to initialize and track separate implicit surfaces that
could then be matched by our algorithm.
\thirdRev{
Additionally, an interesting venue for future work
would be the inclusion of Eulerian advection schemes that propagate
information forward (instead of backward) \cite{lentine2011mass}. These algorithms are typically more complicated,
but could circumvent some of the alignment problems discussed in \myrefsec{sec:alignment}.}

For larger data sets, hard-disk access can become a noticeable
component of the runtimes.  If this becomes a bottleneck, compression
schemes to reduce the size of the stored volumes could be introduced.
Also, our interpolation would need to be changed to guarantee
smooth transitions across simplex boundaries, and
we defer an extension of our interpolation
scheme with bi-directional deformations along simplex edges to future work.
\scndRev{
For interpolations with more than three dimensions, it is also non-trivial
to generate simplicial tessellations. While regular data points could be manually
tessellated for higher dimensions, arbitrary sample locations would be tricky to deal with. }

Lastly, we have ignored the underlying physics during the optimization procedure.
Our intention was to explore how far a purely data-driven optical
flow solve could be taken in this setting.  
However, the algorithm could be readily combined with 
other optimization schemes \cite{Gregson:2014:divFreeof}
to incorporate additional physical constraints. For example, it will be highly
interesting to include constraints for mass conservation of the deformed motion from one frame
to the next. 

\section{Conclusions}
\label{sec:conclusions}

\begin{figure*}[bt] {  \vspaceImgSm
	\begin{center}  \includegraphics[width=0.96\linewidth]{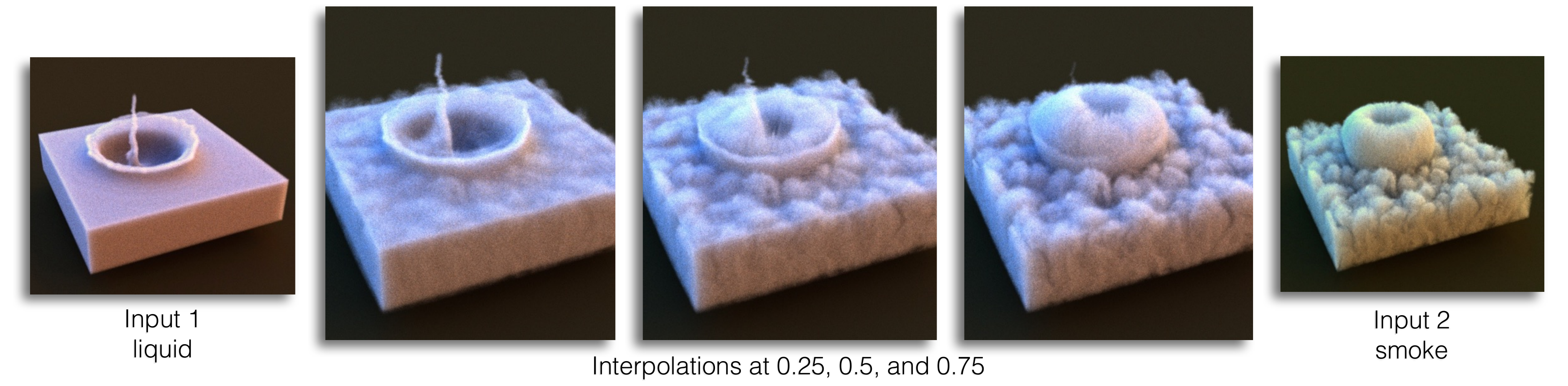}
	\end{center} \vspaceImgSm
} \caption{ \label{fig:liqsmComp} 
	On the far left and right the inputs are shown: a drop of liquid, 
	and a (cold) blob of smoke, each impacting a basin of liquid and smoke, 
	respectively. The different interpolations in the center (with a resolution of $300^3$) illustrate our matching of the two different phenomena.
} \end{figure*}

We have presented an unconventional way to employ optical flow for
interpolating space-time data of fluid simulations.  The resulting algorithm
can find complex 4D registrations without user input, and is able
to match phenomena as different as the smoke plume and liquid drop in
\myreffig{fig:liqsmComp}.

This could be highly interesting not only for fluid animations, but also for other types
of data, such as character animations without temporally coherent meshes (connectivity
information quickly gets lost in the different stages of a production pipeline).
Here, additional constraints such as piecewise rigidity will be interesting avenues
for future work. %
\revision{
We also envision our work to be very useful in the context of
precomputing complex interactive effects \cite{stanton2014srg}, and
we are currently exploring its application in real-time settings. 
The simplicity of the calculations for surface extraction  
should make interactive frame rates on mobile devices possible.
}

Finally, it is an exciting outlook to have tools that automatically
analyze and register large collections of space-time data sets. 
This would open up numerous interesting applications for offline as well as 
real-time effects beyond data-driven fluid simulations. %

\section{Acknowledgements}

This work was funded by the ERC Starting Grant {\em realFlow} (StG-2015-637014). %
Thanks to Karthik Raveendran, Chris Wojtan and Greg Turk
for the interesting discussions about blending and interpolating liquids.
Additionally, thanks to James Gregson, Ivo Ihrke, Wolfgang Heidrich
and Daniel Cremers for insights and discussions about optical flow.
Kudos to Ted Kim and Tiffany Inglis for proof-reading early versions, and
to Rachel Chu for help with the video.
Finally, thanks to the anonymous reviewers for the many helpful suggestions
improving the paper.

\renewcommand{\baselinestretch}{0.995} \normalsize 
\bibliographystyle{acmsiggraph}
\bibliography{ofblendArxiv}

\renewcommand{\baselinestretch}{1.00} \normalsize

\end{document}